\documentstyle[psfig]{mn2e}

\def\gs{\mathrel{\raise0.35ex\hbox{$\scriptstyle >$}\kern-0.6em
\lower0.40ex\hbox{{$\scriptstyle \sim$}}}}
\def\ls{\mathrel{\raise0.35ex\hbox{$\scriptstyle <$}\kern-0.6em
\lower0.40ex\hbox{{$\scriptstyle \sim$}}}}
\def\ls{\mathrel{\hbox{\rlap{\hbox{\lower4pt\hbox{$\sim$}}}\hbox{$<$}}}}
\def\gs{\mathrel{\hbox{\rlap{\hbox{\lower4pt\hbox{$\sim$}}}\hbox{$>$}}}}

\def\mnras {{\sc MNRAS}}
\def\apj {ApJ}
\def\apjs {ApJS}

\def\aj {AJ}

\def\pasp {PASP}

\title[Abell 1386 \& the Sloan Great Wall]
      {The architecture of Abell 1386 and its relationship to the Sloan Great Wall}
\author[K.\,A.\ Pimbblet, et al.]
       {Kevin A.\ Pimbblet\,$^{1}$\thanks{email: Kevin.Pimbblet@monash.edu}, 
	Heinz Andernach\,$^{2,3}$,
      Cherie K.\ Fishlock\,$^{4}$, 
\and Isaac G.\ Roseboom\,$^5$, \& Matthew~S.~Owers\,$^{6}$
        \vspace*{1mm}\\
$^1$School of Physics, Monash University, Clayton, Victoria 3800, Australia\\
$^2$Argelander-Institut f\"ur Astronomie, Universit\"at Bonn, D-53121 Bonn, Germany\\
$^3$Departamento de Astronom\'{i}a, 
Universidad de Guanajuato, AP~144, Guanajuato CP 36000, Mexico\\
$^4$Department of Physics, University of Queensland, Brisbane, Queensland 4072, Australia\\
$^5$Department of Physics and Astronomy, University of Sussex, Falmer, East Sussex BN1 9QH, UK\\
$^6$Centre for Astrophysics and Supercomputing, Swinburne University of Technology, 
Hawthorn, Victoria 3122, Australia\\
}

\date{\fbox{\sc Draft: \today\ --- Do Not Distribute}}
\pagerange{000--000}

\begin{document}

\maketitle

\begin{abstract}
We present new radial velocities from AAOmega on the Anglo-Australian
Telescope for 307 galaxies ($b_J < 19.5$) in the region of the rich
cluster Abell~1386.  Consistent with other studies of galaxy clusters 
that constitute sub-units of superstructures, we find that 
the velocity distribution of A1386 is very broad 
(21,000--42,000\,km\,s$^{-1}$, or $z=0.08$--0.14) and complex.  
The mean redshift of the cluster that Abell designated as number 1386 is 
found to be $\sim0.104$. However, we find that
it consists of various superpositions of line-of-sight components.
We investigate the reality of each component by testing for 
substructure and searching for giant elliptical galaxies in each and show 
that A1386 is made up of at least four significant
clusters or groups along the line
of sight whose global parameters we detail.
Peculiar velocities of brightest galaxies 
for each of the groups are computed and
found to be different from previous works, 
largely due to the complexity of the sky area and the depth of 
analysis performed in the present work.
We also analyse A1386 in the context of its parent 
superclusters: Leo~A, and especially the Sloan Great Wall.
Although the new clusters may be moving toward mass concentrations in 
the Sloan Great Wall or beyond, many 
are most likely not yet physically bound to it.
\end{abstract}

\begin{keywords}
galaxies: clusters: individual: Abell 1386 ---
galaxies: kinematics and dynamics --- 
catalogues
--- large-scale structure of Universe
--- galaxies: elliptical and lenticular, cD
\end{keywords}

\section{Introduction}
Galaxies are typically found clustered together with other
galaxies -- whether this be in small groups, or large
rich galaxy clusters that contain $\sim 10^3$ members.  In turn,
these objects can be clustered together into superclusters
and joined in a complex manner via filaments of galaxies to form
the now familiar web-like or sponge-like structure (Gott et al.\ 1986) seen
in modern redshift surveys (e.g.\ Colless et al.\ 2001).
Amongst the first systematic redshift slice surveys in the 
early 1980s was the CfA2 survey. The survey revealed evidence for
the so-called `Great Wall' (De Lapparent et al.\ 1986; Geller \& Huchra 1989;
Ramella et al.\ 1992).  
This structure was found to extend over 100 degrees in the 
sky, passing 
from A779 in the West and 
through the Coma and A1367 galaxy clusters
up to A2199 at its Eastern end, thus 
having a physical size of $\sim$160\,$h_{75}^{-1}$\,Mpc.
Although more large-scale filaments have been noted in 
the literature since the discovery of the Great Wall (e.g.\ 
Pimbblet, Drinkwater, \& Hawkrigg 2004; 
Bharadwaj et al.\ 2004; Porter \& Raychaudhury 2005), 
the largest known (local) 
structure to date is the Sloan Great Wall (Tegmark et al.\ 2004;
Gott et al.\ 2005; 
Nichol et al.\ 2006;
Einasto et al.\ 2010) 
which is 80 per cent longer than 
the CfA2 Great Wall.  
Finding and refining our knowledge about 
very large structure in the Universe alongside contrasting them with
predictions from a variety of structure formation scenarios is 
highly beneficial to a number of areas of extra-galactic
research ranging from determining the homogeneity scale to 
testing whether our dark matter description of the evolution
and topology of structure in
the Universe is correct (cf.\ Yaryura, Baugh \& Angelo 2010; 
Gott et al.\ 2008; Pimbblet et al.\ 2004; Hara \& Miyoshi 1993; 
Park 1990; White et al.\ 1987).  

Over the past few years, we have been actively compiling redshift data for
over 1000 Abell clusters with the aim of calculating the peculiar velocities
of their brightest cluster members  (BCM; Coziol et al.\ 2009; Pimbblet 2008;
Pimbblet, Roseboom, \& Doyle 2006). 
A1386 drew our attention during this compilation
effort as it was the only cluster in this sample of $\sim1200$ clusters
for which a BCM could be identified whose radial velocity coincided with
one of each of the three sub-units identified 
along the line of sight (Coziol et al.\ 2009).
The lowest-redshift BCM (2MASX~J11481434$-$0159000) turned out to have a 
very high peculiar velocity just above the cluster's radial velocity dispersion
of $\sim1180$\,km\,s$^{-1}$. Thus A1386 was included as one of several target
clusters with BCMs of both low and high peculiar velocities in order 
to study possible relations of cluster substructure with BCM peculiar velocity.
As it happens, A1386 is also a member of the Leo~A supercluster
(SCL100 in Einasto et al.\ 1997; cf. Pimbblet, Edge \& Couch 2005) which itself
is part of the even larger Sloan Great Wall (Gott et al.\ 2005; 
Einasto et al.\ 2010). Therefore
a deep redshift survey of its surroundings appeared to offer new insights
into the structure and depth of such large aggregates of 
galaxy clusters.  

In this paper, we present new observations of A1386 taken with AAOmega on the
Anglo-Australian Telescope.  In Section 2, we describe these observations in
detail, including the galaxy selection and completeness.  We examine the
robustness of our new radial velocities in Section 3.  In Section 4, we present
a full analysis of the dynamics of A1386 and other objects along the line
of sight to provide a better understanding of the state of this unusal cluster.
In Section 5, we consider this cluster in the context of the Sloan Great Wall.
We summarize our findings in Section 6 and present our new radial
velocities in the Appendix.
Throughout this paper we use 
H$_0 = 75$ $h_{75}$\,km\,s$^{-1}$\,Mpc$^{-1}$,
$\Omega_M=0.27$, and $\Omega_{vacuum}=0.73$.

\section{Data and Reduction}
The observations for this work are from AAOmega two degree field (2dF)
multi-fibre spectroscopy at the Anglo-Australian Telescope, Australia.  
AAOmega is the 2006 upgrade to the 2dF spectrograph
(Lewis et al.\ 2002).
Unlike 2dF, AAOmega is a dual-beam spectrograph that is able
to cover a wavelength range of 3700--8500\,\AA. Similar to
its predecessor, AAOmega can achieve the simultaneous observation
of up to $\sim$400 targets (including guide stars) in any single configuration
(see www.aao.gov.au/AAO/2df/aaomega/aaomega.html).

Our observations 
were made in a mixture of conditions.  For our first set of observations
taken on 25 March 2007, thick cloud and fog spoilt the spectra
of all targets.  On the second night, the seeing was large (2.5 arcsec),
but otherwise the conditions were ideal (i.e.\ photometric). 
Here we elect only to use the observations from 26 March 2007 and discard
the weather affected observations.

The targets for our observations are chosen in a similar
manner to Pimbblet et al.\ (2006). 
In brief, we make use of the APM catalogue 
(e.g.\ Maddox et al.\ 1990; see also
www.ast.cam.ac.uk/$\sim$mike/apmcat/) to select all objects
flagged as galaxies in both $b_J$ and $r_F$ passbands
in order to create a sample that will not be highly contaminated
by Galactic stars and not biased with regard to galaxy colour
(i.e.\ we do not just select elliptical galaxies that lay on the 
colour-magnitude red sequence).
The APM positional accuracy is 
better than 0.3 arcsec and is therefore more than sufficient
for AAOmega observations.  Moreover, this approach is 
the same approach used by Colless et al.\ (2001) for the 
Two Degree Field Galaxy Redshift Survey (2dFGRS), 
being more complete for fainter galaxies. Targets were chosen
within a box of $-2.7^\circ<Dec<-1.2^\circ$ 
and $176.3^\circ<RA<177.8^\circ$ (equinox J2000).

In making the AAOmega observing configuration, we assigned a
priority to each target galaxy based upon its $r_F$ magnitude
such that the highest priority is given to the brightest 
galaxies.  This was done not only to obtain the best possible
magnitude-limited sample of galaxies, but also to alleviate any possible effect
from poor weather or fibre positioning, as brighter galaxies
are more likely to generate good-quality spectra.
We did not perform any down-weighting of targets 
when these had literature redshifts.
Guide star candidates were also generated from the APM catalogue 
in the magnitude range $13.75 < r_F < 14.25$ and were 
quality-controlled (by eye)
to ensure that they were isolated.  Blank sky positions were 
provided by the software and
down-selected so that none of them 
were accidentally placed on 
top of `real' objects and 
spoiled our sky subtraction.

\begin{figure}
\centerline{\psfig{file=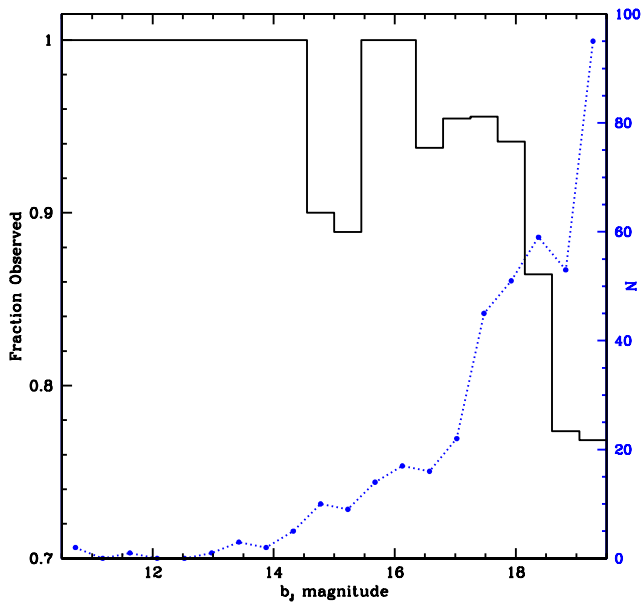,angle=0,width=3.4in}}
%\epsscale{1.10}
%\plotone{magfrac.ps}
  \caption{Fraction of galaxies from the parent APM catalogue
that are selected for observation with 2dF (solid line histogram and 
left hand vertical axis) with the total number of galaxies 
available for selection (dotted line; right hand vertical axis).
}
\label{fig:mag}
\end{figure}

In Fig.~\ref{fig:mag}, we plot all the observed galaxies 
out of the total possible number of potential targets 
as a function of magnitude, down to the limit of $b_J=19.5$.
All of the brighter galaxies ($b_J < 14.5$) were observed successfully,
with the fraction falling to less than 80 per cent at $b_J=19.5$.
The dip in fraction observed 
in the region of $b_J\sim15$ and some of 
the fall toward fainter targets is due to one of two effects:
(a) there is a fibre crossing the target in order to hit
a brighter (i.e.\ higher priority) target; or (b) the
target is in close proximity to another target of equal or
higher priority which would cause a fibre collision should both
targets be observed.  At fainter magnitudes,
there are not enough fibres left to be placed on all possible 
targets and hence many of them simply do not get observed.

%
%  FIGURE.  Fraction observed
%

Our observations used the 580V and 385R gratings which yield a 
central dispersion of 1.0 and 1.6 \AA\,pixel$^{-1}$ respectively.  
These AAOmega spectra are generally superior in quality to the 2dFGRS survey
(Colless et al.\ 2001) given both the higher spectral resolution and
coverage of a wider wavelength range,
and are generally well-able to yield secure redshifts despite
the somewhat inclement observing conditions.

Our dataset was reduced with the 2dF data reduction pipeline
in a standard manner (see www.aao.gov.au/2df/).
This included a Laplacian Edge Detection step to reject cosmic
rays from our data in an efficient manner (see Farage \& Pimbblet 2005
and 
references therein for a full discussion of the benefits of this methodology).
To obtain redshifts of our targets, we made use of the {\sc zcode}
package that was originally employed on 2dFGRS by Colless et al.\ (2001),
and we refer the reader to that publication for more explicit details.
The output of the code consists of a cross-correlation with
the best-matched template spectra, i.e.\ those with the highest $R$
value according to Tonry \& Davis (1979). 
Our template spectra range from G stars, through
globular clusters, right out to galaxy spectra.  
Each redshifted target was then de-redshifted to rest-frame wavelengths
and checked by eye (by KAP) to ensure that the emission and absorption
features are in the correct locations.
The fraction of our targets that produce reliable redshifts is high:
87 per cent of all our targets (100 per cent for $b_J<15.0$, 
only dropping below 80 per cent at $b_J>19.25$).
We present our redshift catalogue in Appendix~A.

\section{Redshifts and Reliability Control}
In total, we obtained redshifts for 307 objects.  
Of these, three are most likely 
stellar in nature as they have velocities of less than 200 km\,s$^{-1}$.
One of them (PRA003) is SDSS~J114948.77$-$014728.2, already suggested to 
be a star in the SDSS database (Abazajian et al.\ 2009), and two others 
(PRA096 and 244) had radial velocities measured in 2dFGRS suggesting 
them to be stars.  

\begin{figure}
\centerline{\psfig{file=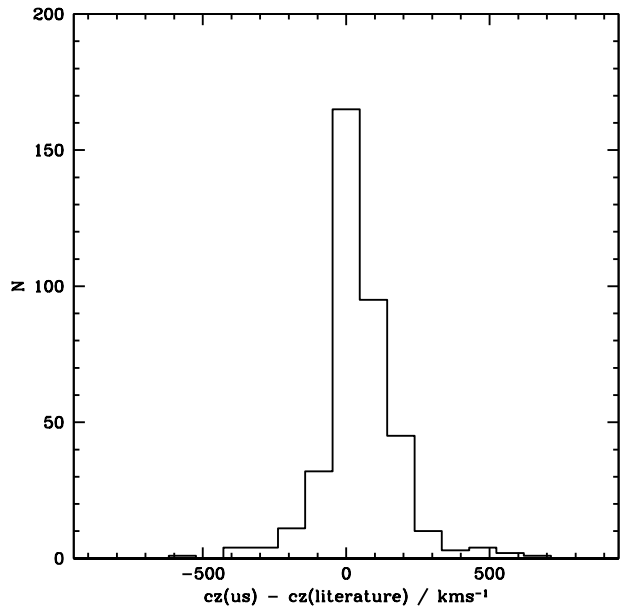,angle=0,width=3.4in}}
%\epsscale{1.10}
%\plotone{diff.ps}
  \caption{Deviation of literature redshifts from our measured redshifts,
including targets that have multiple matches to various different 
sources.  
}
\label{fig:diff}
\end{figure}

To further 
probe the reliability of our redshift measurements we proceed by
comparing our redshifts to those already published in the
literature.  For this comparison, we make use of a number of other
catalogues that possess an overlap with our own
observations: 
2dFGRS (Colless et al.\ 2003); 
2QZ (Croom et al.\ 2004);
6dFGS (Jones et al.\ 2009);
SDSS Data Release 7 (DR7; Abazajian et al.\ 2009); 
Da\,Costa et al.\ (1998);
Doyle et al.\ (2005);
Falco et al.\ (1999);
Grogin et al.\ (1998);
Quintana \& Ram\'\i rez (1995); 
Shectman et al.\ (1996); 
Slinglend et al.\ (1998); \&
Theureau et al.\ (2004).
In some cases, a target in our catalogue appears in several of the
above catalogues, each with a different reported redshift.
In Fig.~\ref{fig:diff}, we plot the difference between our
measured redshifts and the literature redshifts.  
The measured mean and median difference in redshift is
$-$23 km\,s$^{-1}$ and 27~km\,s$^{-1}$, respectively.  
When performing this analysis, we found
two especially note-worthy, low redshift, 
discrepancies between our redshifts 
and previously published ones, 
which exceed by far the quoted redshift errors.
The first major difference
is between ourselves and 6dFGS (PRA284; Appendix~A) 
-- this is due to a poor quality 6dFGS spectrum (D.H.~Jones, priv.~comm.).

%
%  FIGURE.  Redshift comparison
%

The second is PRA037 (also 2MASX~J11481434$-$0159000; Appendix~A)
when compared to 
Quintana \& Ram\'\i rez (1995). This galaxy also has 
published redshifts in three other catalogues.
Whilst our redshift is 
non-discrepant with both 2dFGRS (Colless et al.\ 2003) 
and 6dF (Jones et al.\ 2009), 
it is $2.6\sigma$ away from that published in SDSS-DR2 (Abazajian et al.\ 2004).
The reason for the discrepancy becomes more obvious from an examination of
this object's image at optical (i.e.\ the Digitized Sky Survey and SDSS)
and NIR (i.e.\ 2MASS) wavelengths: it possesses a secondary core
or neighbour galaxy 6 arcsec away at PA~$\sim230^{\circ}$.  We therefore
contend Quintana \& Ram\'\i rez (1995) on the one hand, and SDSS-DR2,
2dFGRS, and 6dF on the other, each measured a different core's velocity.

\section{Analysis}

\subsection{Dynamics}
\begin{figure}
\centerline{\psfig{file=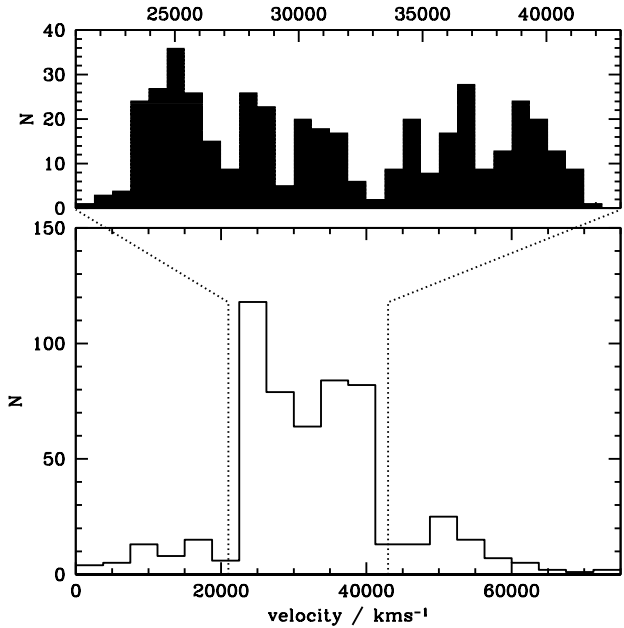,angle=0,width=3.4in}}
%\epsscale{1.10}
%\plotone{zhist.ps}
  \caption{Velocity histogram for our observations combined 
with further velocities obtained from the literature. The upper 
panel of the figure displays an enlarged region from the lower
panel.  The velocity structure of field of 
A1386 is complex and likely comprises at least four sub-clusters or groups.
}
\label{fig:zhist}
\end{figure}

We begin our analysis of the dynamics and architecture of A1386 by
examining its velocity structure. In constructing the velocity histogram
(Fig.~\ref{fig:zhist}), we include not only the objects from our
new observations, but also all objects 
with redshifts available from the literature (see Section 3, above, 
and Appendix).  
Measurements that belong to the same galaxy were identified, based on their 
close positional coincidence, and the velocity with smallest error was chosen
for that galaxy.

%
%  FIGURE.  Redshift histogram
%

The velocity distribution (Fig.~\ref{fig:zhist}) is
unusually broad, with several sub-peaks,
and there appears to be rich and highly complex substructure
in the core of A1386.  The complexity is perhaps not unexpected 
given that A1386 resides within the supercluster Leo~A 
(SCL100 in Einasto et al.\ 1997; cf. Pimbblet, Edge \& Couch 2005) 
and hence 
is a part of the Sloan Great Wall (see Fig.~9 of Gott et al.\ 2005).
We note that this broad peak
in velocity is, however, distinct and isolated from
other foreground and background structures. Indeed, this cluster
is very isolated in redshift space: 
the closest cluster in redshift space is WBL~355 at $z\sim0.028$ 
(White et al.\ 1999), some 52~arcmin due WNW from the centre of A1386.
We refrain from making further analysis
of WBL~355 as our observations are only just probing the outskirts
of this poor cluster.

We note that the redshift of A1386 ($z=0.1018$) given 
by Struble \& Rood (1999)
is based on that of a single galaxy 
(published by Quintana \& Ram{\'{\i}}rez 1995)
whose redshift 
is coincidentally located near the middle of the velocity distribution.
However, it presents a real problem in trying to compute 
any `mean' redshift of the population, let alone a meaningful
velocity dispersion (cf.\ Abell~779 in Oegerle \& Hill 2001).
This can readily be illustrated by a simple application of 
the $3\sigma_v$ clipping technique of Yahil \& Vidal (1977) where
the mean velocity and dispersion of a cluster are determined by
iteratively clipping any galaxy that is greater than $3\sigma_v$
from the mean of the velocity distribution.
Using an initial clip of 21000\,km\,s$^{-1}  < cz < 42000$\,km\,s$^{-1}$, we 
obtain a mean velocity of $31241\pm275$\,km\,s$^{-1}$ and
an unphysically large (and clearly erroneous) velocity dispersion 
of 5723~km\,s$^{-1}$. We are thus forced to split up
the cluster into more sensibly sized sub-components.
Fig.~\ref{fig:zhist} would suggest that there are multiple
sub-peaks in the overall velocity distribution and therefore we 
proceed with the aim of attempting to isolate these peaks.

\subsection{Substructure}
Based on an inspection of Fig.~\ref{fig:zhist}
it is likely that A1386 has \emph{at least} four components 
(mean velocities of approximately 25000\,km\,s$^{-1}$, 28500\,km\,s$^{-1}$, 
31000\,km\,s$^{-1}$ and 37000\,km\,s$^{-1}$; see Fig.~\ref{fig:zhist})
that make up what Abell (1958) originally defined as the cluster proper.
In order to better delineate the structure of the cluster,
we now apply the Dressler \& Shectman (1988; DS) test to our 
catalogue -- one of the most sensitive 
general tests for substructure available
(Pinkney et al.\ 1996; see also Section 4.4).  
For each cluster member,
the DS algorithm computes the mean 
local velocity, $\overline{cz}_{local}$,
and local standard deviation, $\sigma_{local}$, of that member's
$N_{local}$ nearest neighbours in projection.  These localized values are
then compared to the global values of the cluster mean velocity,
$\overline{cz}$, and
cluster velocity standard deviation, $\sigma_v$,
to produce a measure of deviation:

\begin{equation}
\delta^2 = \left( \frac{ N_{local} + 1 }{\sigma_v^2} \right) 
[ (\overline{cz}_{local}-\overline{cz})^2 + ( \sigma_{local}-\sigma_v)^2 ] 
\end{equation}
that can be utilized to locate clumps of spatially close deviant galaxies.
Consistent with DS, in this work we use 
the 10 nearest neighbours (i.e.\ $N_{local} + 1 = 11$) to compute $\delta$.
A cumulative quantity, $\Delta$, is then found by summing all values
of $\delta$ for the cluster.  By comparing $\Delta$ to Monte Carlo 
simulations in which the member's velocities are shuffled around the
positions, we can estimate a confidence level for the overall probability
of substructure in the cluster.

\begin{figure*}
\centerline{\psfig{file=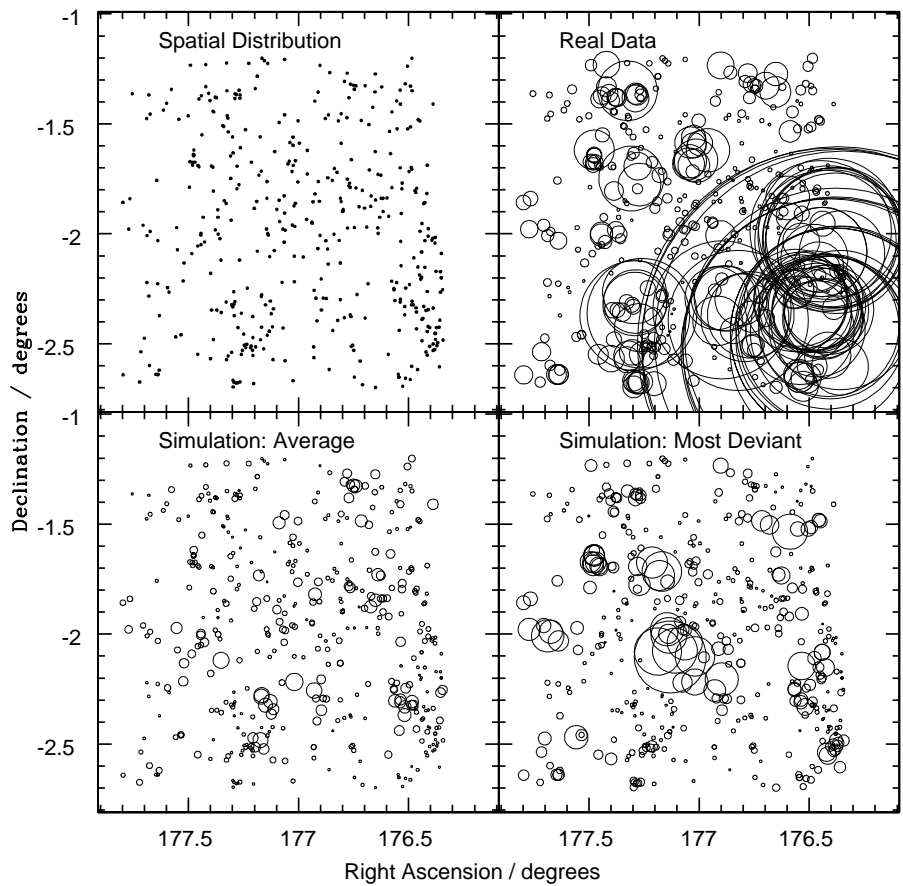,angle=0,width=5in}}
%\epsscale{1.10}
%\plotone{dstest.ps}
  \caption{Results of the DS test.  
Top left: spatial distribution of all galaxies with $21000$~km s$^{-1} < cz < 42000$~km s$^{-1}$ .
Top right: Result of the DS test as applied to A1386 -- 
spatially close overlapping
circles denote likely substructure locations.  
Bottom left: average of 1000 Monte Carlo simulations.  
Bottom right: the single most deviant of 1000 Monte Carlo simulations.
}
\label{fig:dstest}
\end{figure*}

We acknowledge that we can already guess that
the DS test will show that the cluster has substructure -- the
primary purpose of applying this technique, however, is to
specify the sky positions of likely sub-components within the cluster.
Figure~\ref{fig:dstest} displays the results of applying the DS test
to our data -- each circle is drawn with a diameter proportional
to the deviation of a given galaxy from the global mean velocity
(here, assumed to be 31241\,km\,s$^{-1}$), 
$e^{\delta}$ (see DS for full details of the test); hence substructure
is interpreted as (spatially close) overlapping circles.
We also display the results of the 
average and most deviant of 1000 Monte Carlo simulations 
in Fig.~\ref{fig:dstest}.

%
%  FIGURE.  DS Test
%

Unsurprisingly, the DS test gives unequivocal evidence for sub-clustering as 
the probability of obtaining the cumulative deviation
found for the cluster 
is very low in comparison to the simulations:
$P(\Delta)\ll 0.001$.

\begin{figure*}
\centerline{
\psfig{file=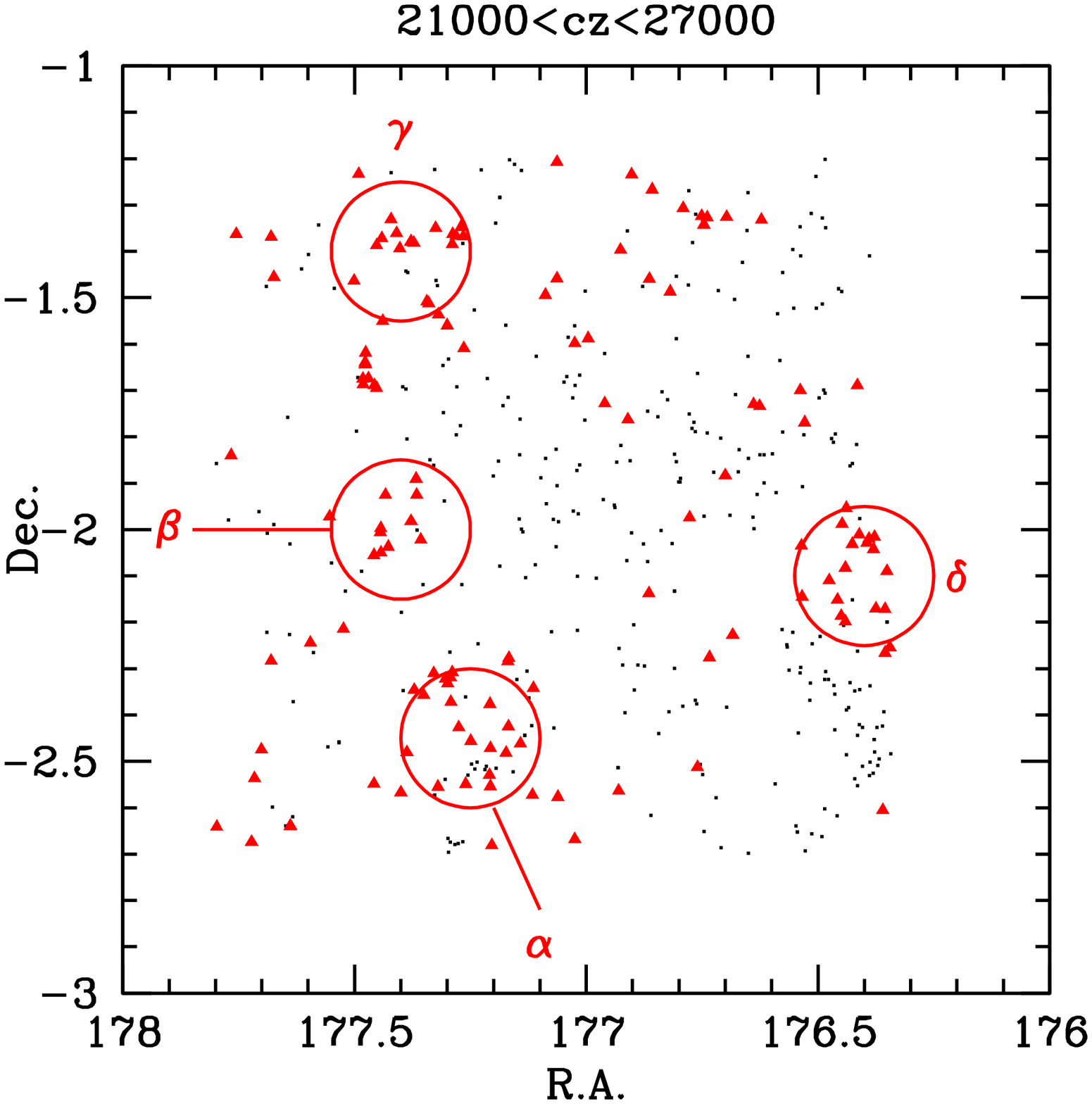,angle=0,width=2.2in}
\hspace*{0.3in}
\psfig{file=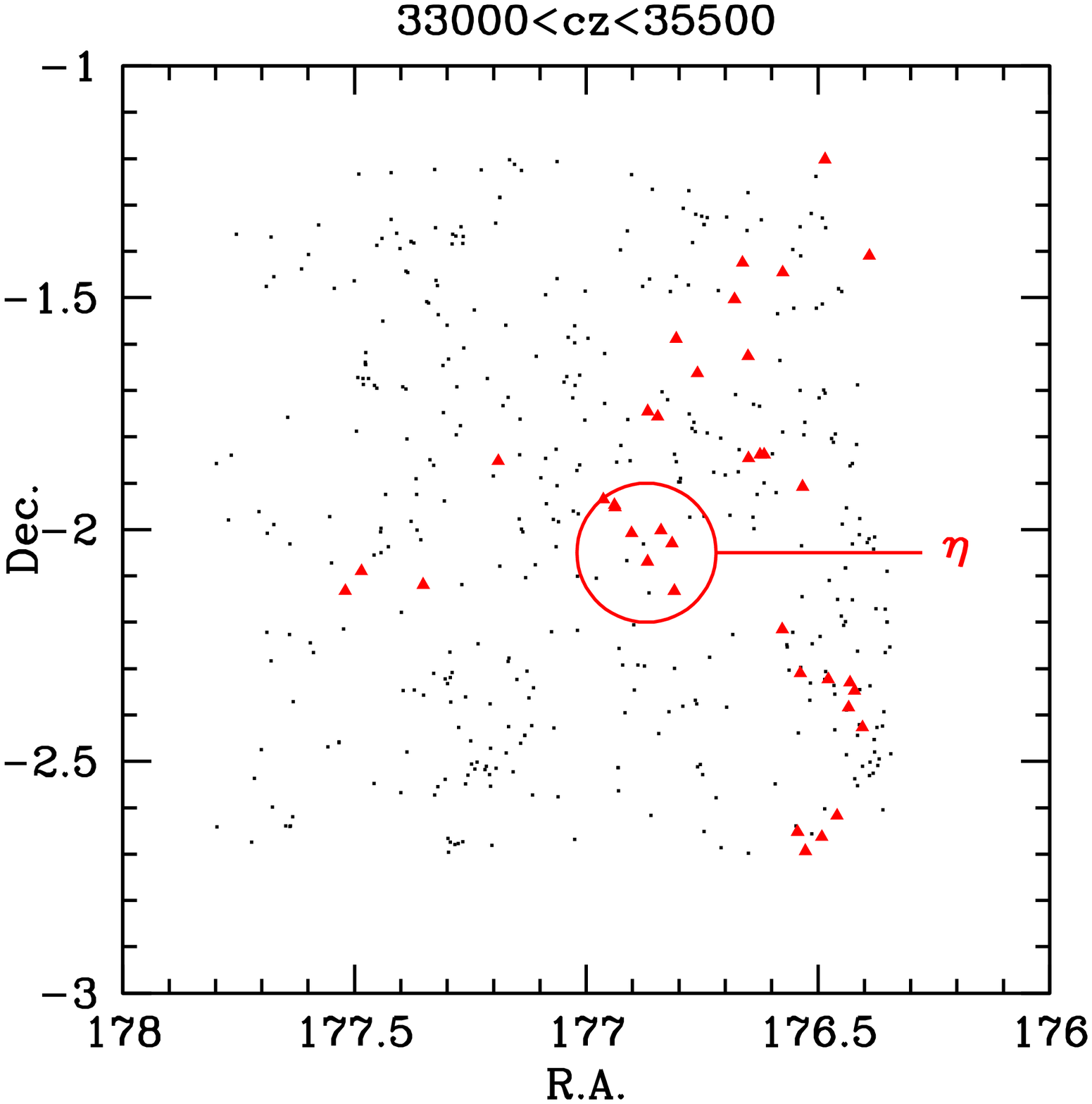,angle=0,width=2.2in}
}
\centerline{
\psfig{file=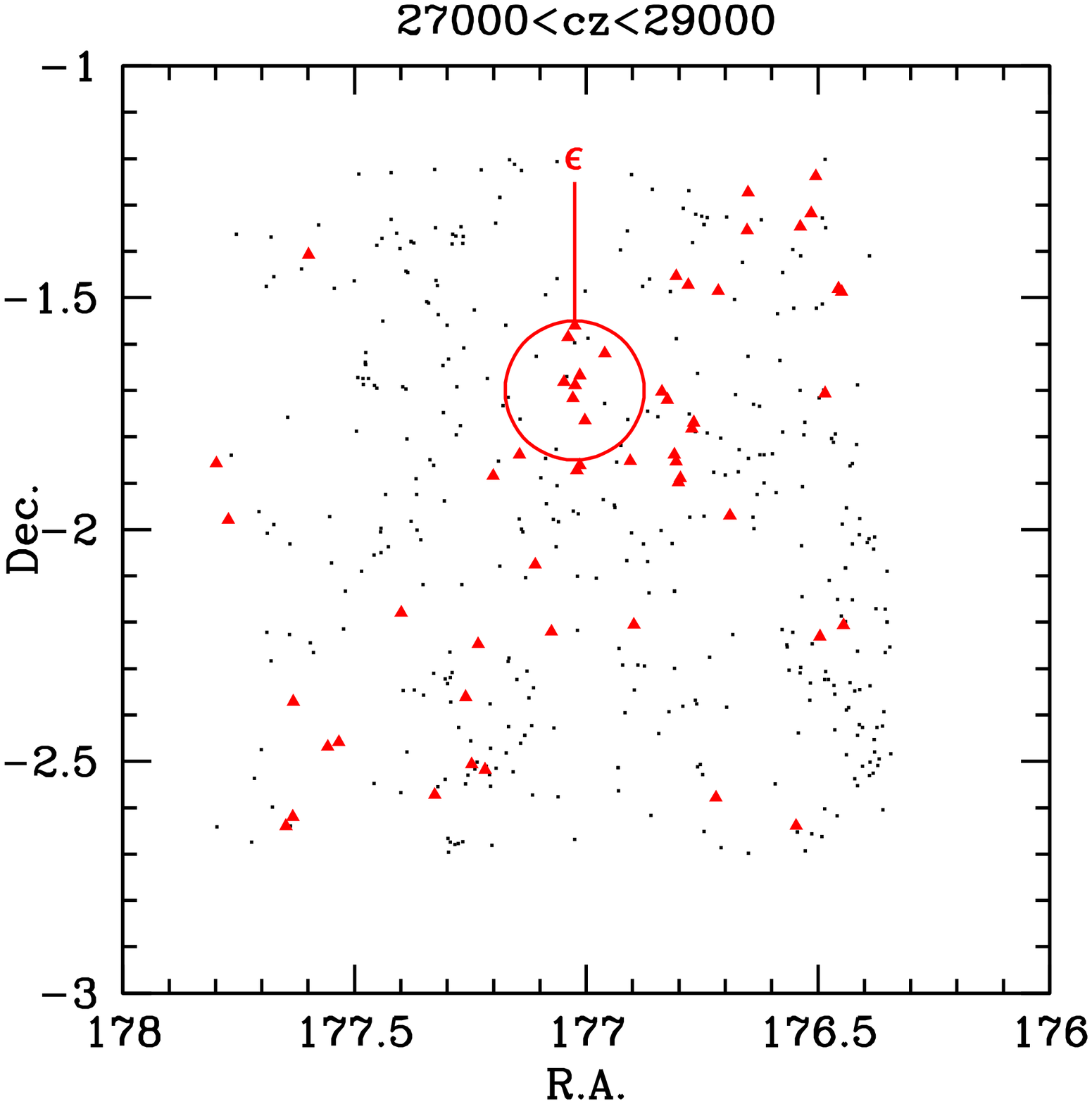,angle=0,width=2.2in}
\hspace*{0.3in}
\psfig{file=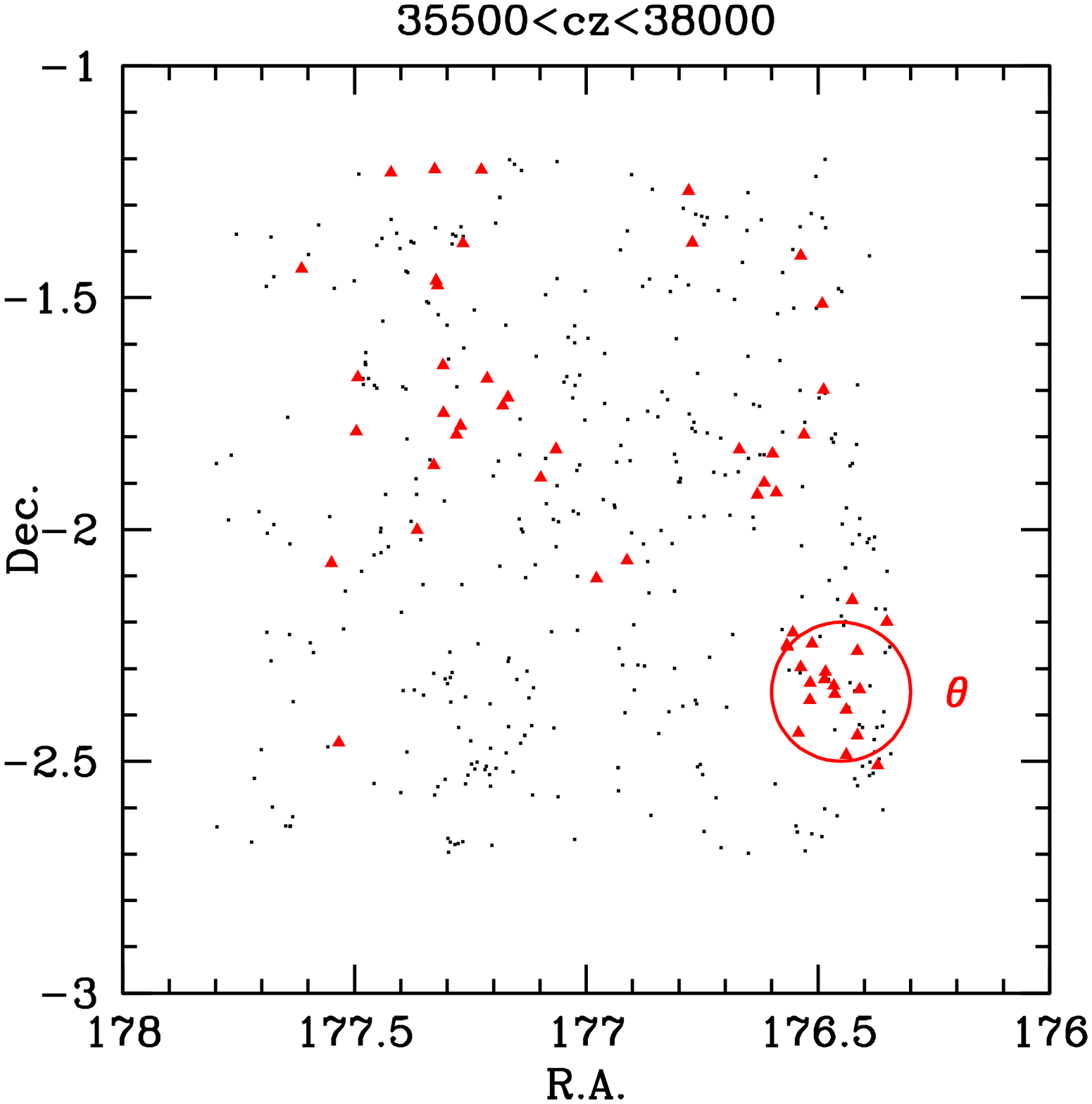,angle=0,width=2.2in}
}
\centerline{
\psfig{file=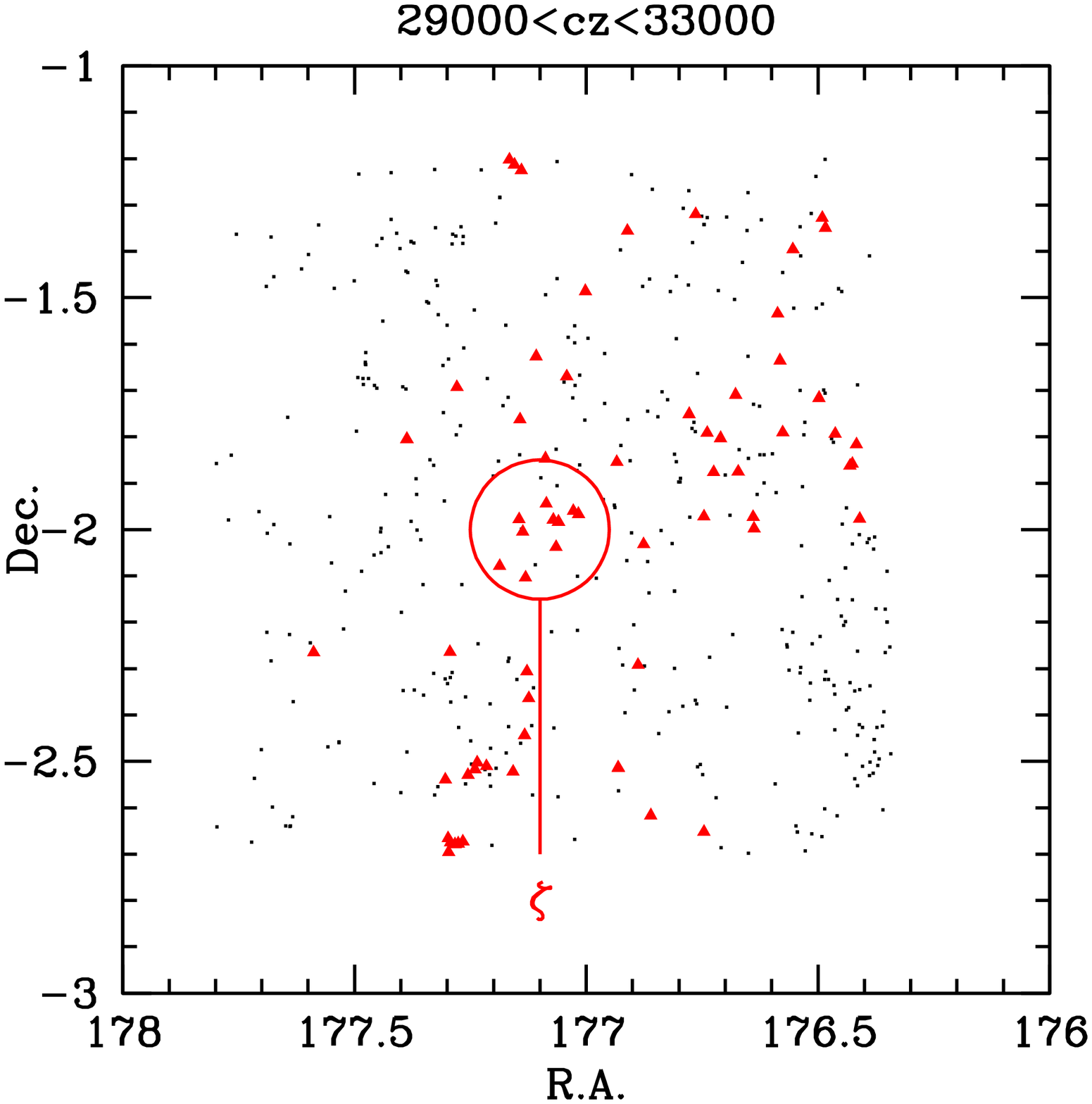,angle=0,width=2.2in}
\hspace*{0.3in}
\psfig{file=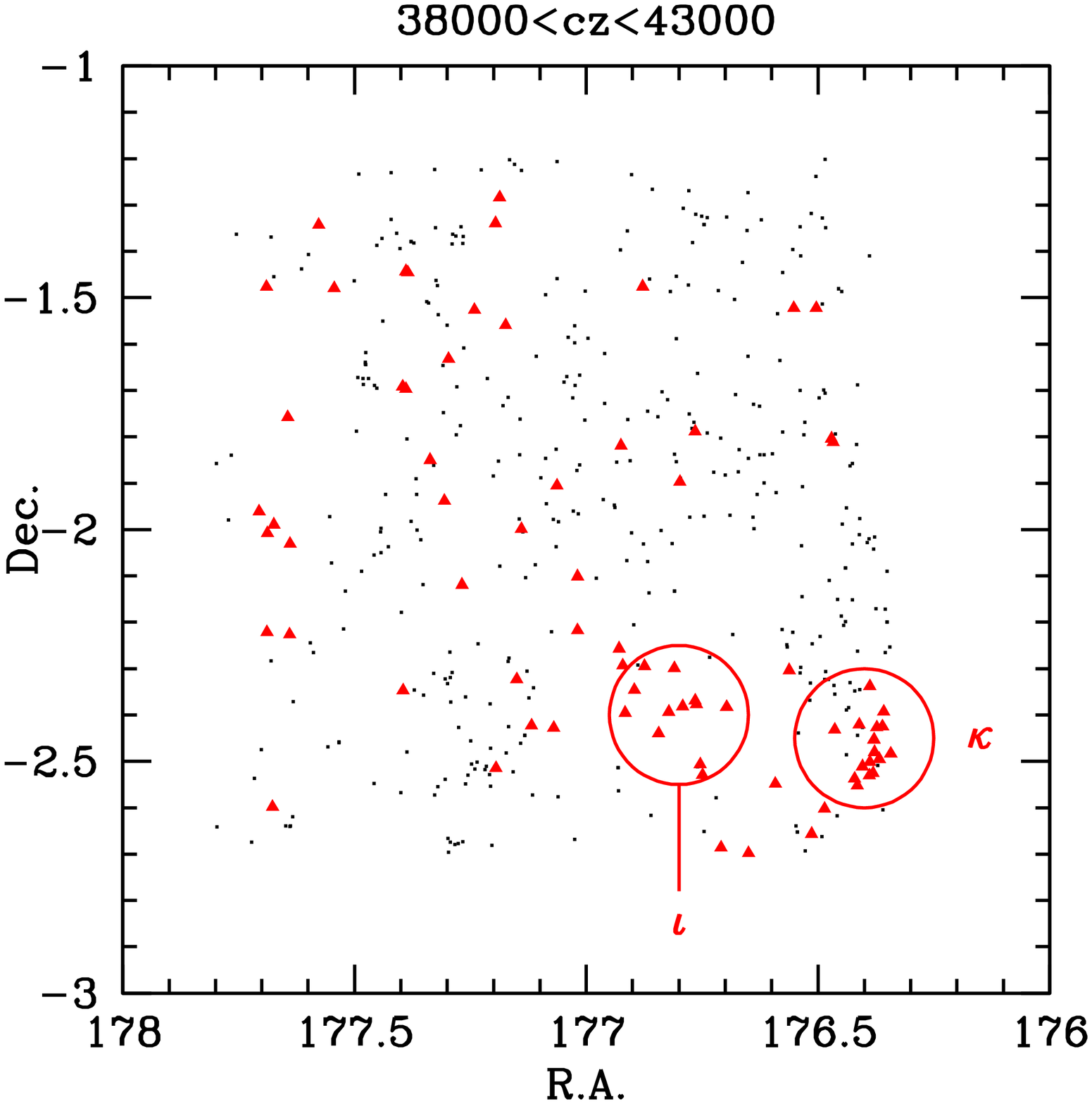,angle=0,width=2.2in}
}
  \caption{Spatial distribution of galaxies in redshift channels
corresponding to the peaks seen in Fig.~\ref{fig:zhist} (triangles);
all other galaxies in the range 
$21000$~km s$^{-1} < cz < 42000$~km s$^{-1}$ are plotted as dots.
Localized overdensities -- i.e.\ likely cluster sub-components -- 
are circled and labelled $\alpha$ through $\kappa$.
}
\label{fig:groups}
\end{figure*}

Interestingly, Fig.~\ref{fig:dstest} displays several regions of
overlapping circles in the real data suggesting localized 
sub-clustering (Dressler \& Shectman 1988).  
We now ask if any of these regions (or, indeed, any localized 
regions at all) correspond to any of the
individual peaks seen in redshift space (Fig.~\ref{fig:zhist}).
To do this, we split the cluster catalogue up into six redshift 
channels that encompass each of the major peaks seen in Fig.~\ref{fig:zhist}
and search (by eye)
for any obvious spatial overdensities.  
The result of this
search is depicted in Fig.~\ref{fig:groups}.

%
%  FIGURE.  subplots of redshift channels
%

We believe that the sub-component marked $\zeta$ is the cluster proper 
because it is nearest to what Abell (1958) catalogued as the cluster centre.  
Comparing Fig.~\ref{fig:groups} to Fig.~\ref{fig:dstest} 
shows that a number of the other sub-components 
(e.g.\ $\delta$)
probably constitute sub-clusters that are
interacting in a complex manner 
with $\zeta$, and potentially with each other as well.  

Of the other overdensities, we suggest that $\theta$ and $\kappa$
are the same entity that extends over two of the redshift channels
in Fig.~\ref{fig:groups}. We identify $\theta$ and $\kappa$ 
as Abell~1373.  By limiting the redshift distribution to the
range $35500<cz<42000$kms$^{-1}$ and localizing the spatial extent
to $176.3^\circ<RA<176.6^\circ$ and $-2.6^\circ<Dec<-2.3^\circ$, 
we compute that 
Abell~1373 has a mean velocity of $38063\pm256$ kms$^{-1}$ and
a velocity dispersion of $1428^{+228}_{-154}$ kms$^{-1}$ 
from 31 members.  
Our values for $\overline{cz}$ and 
$\sigma_v$ are comparable (i.e.\ within $3\sigma$) to those computed
in the recent study of rotating galaxy clusters by
Hwang \& Lee (2007) despite using a different spatial extent.

We have also searched the NASA/IPAC Extragalactic Database (NED,
nedwww.ipac.caltech.edu) for possible clusters that may correspond to
overdensities $\alpha$ through $\kappa$ in order to check whether they
were known before. We found reasonable matches 
with clusters reported by 
Estrada et al.\ 2007; Koester et al.\ 2007; Miller et al.\ 2005; 
and Merch{\'a}n \& Zandivarez 2002, 
and present these in Table~\ref{tab:search}, along
with basic data on the overdensities themselves.  

We have approximated
the mean velocity of each group by taking all galaxies within the marked
circles (radius of 0.15$^\circ$ on the sky; equivalently 1 Mpc at $z=0.1$) 
in Fig~\ref{fig:groups} for the 
purpose of matching them to the literature.  We will refine these
values later in this work by undertaking detailed decomposition
work using Kaye's Mixture Model.  We note that a number of our groups
have reasonable matches to known literature clusters ($\alpha$, 
$\delta$, $\zeta$, $\theta$ \& $\kappa$).  
A further two matches ($\epsilon$ \& $\eta$) 
are found within the MaxBCG catalogue of
Koester et al.\ (2007).  Although the 
MaxBCG photometric cluster redshifts are not
quite the same as our spectroscopic cluster redshifts, 
they are of the order of the quoted photometric
redshift error of $\Delta_Z=0.01$ given by Koester et al.\ (2007)
away from our estimates.
Therefore we regard these matches as plausible.
Finally, $\theta$ is 
identified as matching EAD2007 236 (Estrada et al.\ 2007) which
is also based on a photometric redshift.  
We suggest that EAD2007 236 may be a foreground extension to A1373. 

%
% TABLE.   Cluster Searches
%
\begin{table*}
\begin{center}
\caption{Known clusters in the literature with close proximity to 
the overdensities identified in Fig~\ref{fig:groups}.  The mean velocity 
quoted for each group is computed from galaxies within the marked circles
(radius of 0.15$^\circ$) in Fig~\ref{fig:groups}.  \hfil}
\begin{tabular}{llllllll}
\noalign{\medskip}
\hline
Group     & RA (J2000)& Dec (J2000)& $\overline{cz} /$\,km\,s$^{-1}$ & NED Match    & RA(NED) & Dec(NED) & cz(NED) \\
\hline
$\alpha$  & 11 49 00 & $-$02 22 36 & 26083                           & SDSS-C4 1121 & 11 48 51 & $-$02 30 37 & 26235 \\
$\beta$   & 11 49 38 & $-$01 59 24 & 23675                           & No plausible match found \\
$\gamma$  & 11 49 36 & $-$01 24 00 & 25177                           & No plausible match found \\
$\delta$  & 11 45 36 & $-$02 06 00 & 24343                           & MZ 07066     & 11 45 29 & $-$02 06 24 & 23769 \\ 
$\epsilon$ & 11 48 05 & $-$01 42 00 & 28243                          & MaxBCG J177.02469-01.68868 & 11 48 06 & $-$01 41 19 & 32393$^\dagger$ \\
$\zeta$   & 11 48 22 & $-$02 00 00 & 30813                           & Abell~1386   & 11 48 22 & $-$01 56 41 & 30519 \\
$\eta$    & 11 47 31 & $-$02 01 48 & 34408                           & MaxBCG J176.79744-01.88906$^\ddagger$ & 11 47 11 & $-$01 53 21 & 30774$^\dagger$ \\
$\theta$  & 11 45 46 & $-$02 19 48 & 36326                           & [EAD2007] 236 & 11 45 52 & $-$02 20 11 & 34704$^\dagger$ \\
$\iota$   & 11 47 12 & $-$02 24 00 & 38975                           & No plausible match found \\    
$\kappa$  & 11 45 31 & $-$02 28 12 & 39092                           & Abell~1373   & 11 45 28 & $-$02 23 40 & 39393 \\
\hline
\multispan{6}{$^\dagger$ Denotes a photometric cluster redshift estimate. \hfil}\\
\multispan{6}{$^\ddagger$ It is possible that $\eta$ is a SW extension of $\epsilon$. \hfil}\\
\noalign{\smallskip}
\end{tabular}
  \label{tab:search}
\end{center}
\end{table*}

\subsection{Brightest Cluster Members}
If the sub-components identified in Fig.~\ref{fig:groups} truly
are sub-clusters or groups, 
then when we examine each separately, they may visually resemble 
such a group.  
For instance, each sub-component may contain a brightest cluster
member (BCM) that is a giant elliptical or a cD class galaxy
with an extended diffuse halo; 
or we may find several galaxies in the act of merging to form such 
a galaxy surrounded by an overdensity of early-type galaxies
(cf. Bautz \& Morgan 1970). Conversely, not finding a cD class
galaxy does not immediately mean that we have not located a cluster.
The majority of Abell clusters are 
of ``late'' Bautz-Morgan type, i.e.\ they lack an obvious central, 
bright and dominant
early-type galaxy; here, we are aiming to build up a
body of evidence for the most probable sub-groups.

\begin{figure*}
\centerline{\psfig{file=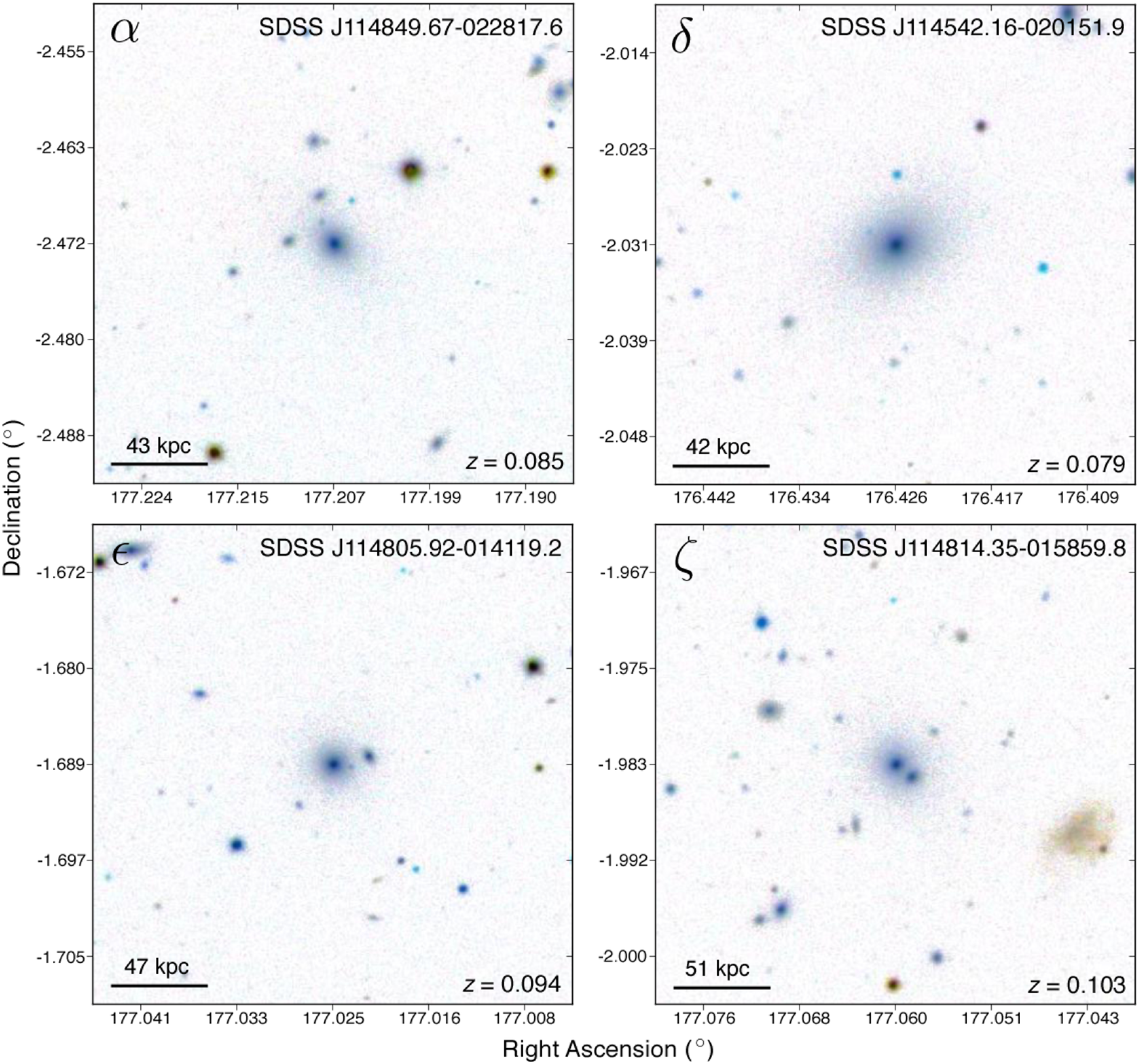,angle=0,width=6.5in}}
%\epsscale{1.10}
%\plotone{subplots.ps}
  \caption{BCMs in the groups identified in Fig.~\ref{fig:groups}.  
All of them have the correct redshifts to be associated with the
groups that they were identified from.
}
\label{fig:adez}
\end{figure*}

In order to perform this test, we inspect SDSS images of the regions centred 
on the sub-components.
Out of the components identified in 
Fig.~\ref{fig:groups}, we find that groups $\alpha$, $\gamma$,
$\delta$, $\epsilon$ and $\zeta$
have obvious bright ellipticals within the regions indicated
in Fig.~\ref{fig:groups}.  Of these, $\alpha$, $\delta$,  
$\epsilon$ and $\zeta$ are all
at the correct redshift but the bright elliptical near the $\gamma$ 
spatial overdensity is a foreground object.  We display these four
BCMs in Fig.~\ref{fig:adez}.  

%
%  FIGURE.  BCM images
%

The BCM in $\alpha$ is SDSS J114849.67$-$022817.6 and was not 
observed in our sample due to fibre collisions and 
prioritization at the field configuration stage (the same is
true for $\delta$ and $\epsilon$).  The 
redshift of this galaxy from SDSS is $cz=25428$ km s$^{-1}$.
We identify the BCMs in $\delta$ and $\epsilon$ as 
SDSS J114542.16$-$020151.9 ($cz=23624$ km s$^{-1}$) and
SDSS J114805.92$-$014119.2 ($cz=28181$ km s$^{-1}$) respectively.
Meanwhile, PRA037 is identified as the BCM of $\zeta$, and 
SDSS J114814.35$-$015859.8 
(also labeled 2MASX J11481434-0159000), with $cz=31022$ km s$^{-1}$ from 
our catalogue.

\subsection{Results of the KMM algorithm}
The DS test is one of the best tests available in a generic three
dimensional case for finding substructure (Pinkney et al.\ 1996).
It is not, however, without its own problems.  Although it
can be readily sensitive from equal mass mergers down to a 
3:1 mass merger ratio given
modest numbers of redshifts (cf. Pinkney et al.\ 1996; Pimbblet 2008),
the number of false positive detections can pose problems.  
Indeed, it \emph{may} be that 
some of the sub-components outlined above do not represent 
true infalling groups or sub-clusters (e.g.\ $\beta$; see 
Fig.~\ref{fig:groups}) given their lack of bright ellipticals.
Moreover, false positive detections can be 
particularly evident for clusters that have features
such as significant radial gradients
in their velocity dispersion profiles (Pinkney et al.\ 1996). With a
structure such as A1386, the likelihood of false positives may be
comparatively high.

To proceed further with delineating the possible sub-components, we
now apply Kaye's mixture model (KMM) algorithm 
to the velocity distribution.  
The KMM algorithm is described in detail in Ashman et 
al.\ (1994) and has been used extensively in the literature (e.g.\ recent
examples include Owers et al.\ 2009; Johnston-Hollitt et al.\ 2008) 
for just this purpose.
In brief, based on a user-supplied number of Gaussians with some initial 
best guess of their central position, the KMM algorithm 
effectively partitions the data into a number of sets and
evaluates whether the fit that results from these Gaussians
is superior to a single Gaussian fit. In all cases studied in this
work, multiple Gaussians are \emph{always} found to be superior fits to
the data than a single Gaussian by the KMM algorithm.  
The key 
questions here are what input should the KMM be given: 
how many Gaussians should one fit to the data and with what initial 
parameters?

In the one-dimensional case, we elect to try two sets of parameters:
the first with four Gaussians and the second with six.  The reason for
these choices stems from visual inspection of Fig.~\ref{fig:zhist}
as discussed above.  For each peak in the
velocity distribution, we use an approximate guess of the 
velocity dispersion ($\sigma_v$) from a visual 
inspection of Fig.~\ref{fig:zhist}.  
We present the input and output parameters for
these two scenarios in Table~\ref{tab:kmm1}.  

We conducted several runs of the KMM algorithm to check how sensitive
the results are to the initial conditions imposed by guessing the
Gaussian's parameters (i.e.\ $\sigma_v$ and $\overline{cz}$).  
This was done by perturbing both $\sigma_v$ and $\overline{cz}$
by incremental amounts.
In the
cases where the perturbation is modest 
($\Delta(\overline{cz})<500$\,km\,s$^{-1}$  and $\Delta\sigma_v<0.5\sigma_v$), 
the KMM algorithm converged
on the same solution.  Therefore, despite guessing the initial
inputs, we regard the output of the KMM algorithm to be
reasonably robust.

From Table~\ref{tab:kmm1}, it appears that a four-Gaussian approach
is incorrect.  The highest-redshift Gaussian (labelled group 4
in Table~\ref{tab:kmm1}) of the four-Gaussian approach has a very
large velocity dispersion of $\sigma_v = 2316$\,km\,s$^{-1}$.  
We regard this as erroneous since the implied cluster mass would be 
unphysically high. However, the first
three Gaussians are good fits to the data and are highly 
plausible.

The six-Gaussian solution appears to be superior to the four-Gaussian
solution at first glance (Table~\ref{tab:kmm1}).  Effectively,
what was group 4 in the four-Gaussian solution has been segmented 
into three new sub-components.  Of these, group 4 looks to
be nearly perfectly partitioned, with groups 5 and 6 both possessing
more reasonable parameters than before.  Arguably, we could
also partition group 6 into two further sub-components (centred
on $\approx$ 38000 and $\approx$ 40000\,km\,s$^{-1}$).  But doing
so results in very few galaxies ($N_{gal}=19$) in one partition and a much reduced
correct allocation rate (76 per cent) indicated by the KMM algorithm.  
Hence we do not regard seven Gaussians as an improvement over the 
six-Gaussian solution in one dimension.

%
% TABLE.   KMM tests
%
\begin{table*}
\begin{center}
\caption{Input and output parameters for the one-dimensional application 
of the KMM algorithm. The table is split into two halves:
the upper half shows the result of inputting four Gaussians to the
KMM algorithm, whereas the bottom half shows the result of 
six Gaussians. The final column shows an estimate of the correct
allocation rate by KMM for each partition.  \hfil}
\begin{tabular}{lll|llll}
\noalign{\medskip}
\hline
      & \multispan{2}{Input Parameters \hfil} & \multispan{4}{KMM Output \hfil} \\
Group & $\overline{cz} /$\,km\,s$^{-1}$ &  $\sigma_v /$\,km\,s$^{-1}$ & $\overline{cz} /$\,km\,s$^{-1}$  &  $\sigma_v /$ km\,s$^{-1}$ & $N_{gal}$ & Rate ($\%$) \\
\hline
1 & 25000 & 1000 & 24807 & 1161 &138 & 98 \\
2 & 28500 & 500 & 28219 & 451 & 56 & 99 \\
3 & 31000 & 750 & 30856 & 818 & 66 & 98 \\
4 & 37000 & 2000 & 37398 & 2316 & 173 & 98 \\
\hline
      & \multispan{2}{Input Parameters \hfil} & \multispan{4}{KMM Output \hfil} \\
Group & $\overline{cz} /$ km\,s$^{-1}$ &  $\sigma_v /$ km\,s$^{-1}$  & $\overline{cz} /$ km\,s$^{-1}$  &  $\sigma_v /$ km\,s$^{-1}$ & $N_{gal}$ & Rate ($\%$) \\
\hline
1 & 25000 & 1000 & 24807 & 1161 & 138 & 98 \\
2 & 28500 & 500 & 28219 & 451 & 56 & 99 \\
3 & 31000 & 750 & 30856 & 818 & 66 & 98 \\
4 & 34500 & 500 & 34507 & 548 & 39 & 98 \\
5 & 37000 & 500 & 36560 & 452 & 49 & 97 \\
6 & 39500 & 1000 & 39442 & 1022 & 85 & 98 \\
\hline
\noalign{\smallskip}
\end{tabular}
  \label{tab:kmm1}
\end{center}
\end{table*}

\subsection{Interpretation}
Taken together, the above suggests 
that the galaxy populations of these other clusters have biased the
cluster and richness identification made by Abell (1958).
There are at least three significant, bona-fide sub-clusters
in the field of A1386: 
these are the first three redshift slices 
specified in Table~\ref{tab:kmm1} (both the four and six
Gaussian KMM solutions).

Of these, the second and third groupings are fairly
solid detections with a single BCM each ($\epsilon$ and $\zeta$;
Figs.~\ref{fig:groups} and~\ref{fig:adez}).  The first 
cluster in Table~\ref{tab:kmm1} 
merits further attention given it has two 
sub-clusters ($\alpha$ and $\delta$; Fig.~\ref{fig:adez})
with potential BCMs (Fig.~6),
both of which have matches to literature clusters (Table~\ref{tab:search}).
On the face of it, this first sub-cluster appears to have
a velocity dispersion ($\sigma_v = 1088$ km s$^{-1}$) that
is typical of a rich, massive galaxy cluster by itself.  
But given the two BCMs and their spatial separation 
(Fig.~\ref{fig:groups}; 
$\approx 2.5$ Abell radii apart), it may be the case that this
redshift grouping is a cluster that is undergoing the early
to mid-stages of a merger event.  

To test this hypothesis, we apply a further DS test, but limit
ourselves to only those galaxies contained in the first 
redshift grouping in Table~\ref{tab:kmm1}.  Given we 
have 138 galaxies in this redshift slice, we will be
sensitive to about a 4:1 mass merger ratio (see Pinkney et al.\ 1996).
The DS test for this redshift slice 
generates a result of $P(\Delta) \ll 0.001$ which 
strongly suggests the presence of substructure.

%
%  FIGURE.  DS test of the first redshift channel
%
\begin{figure}
\centerline{\psfig{file=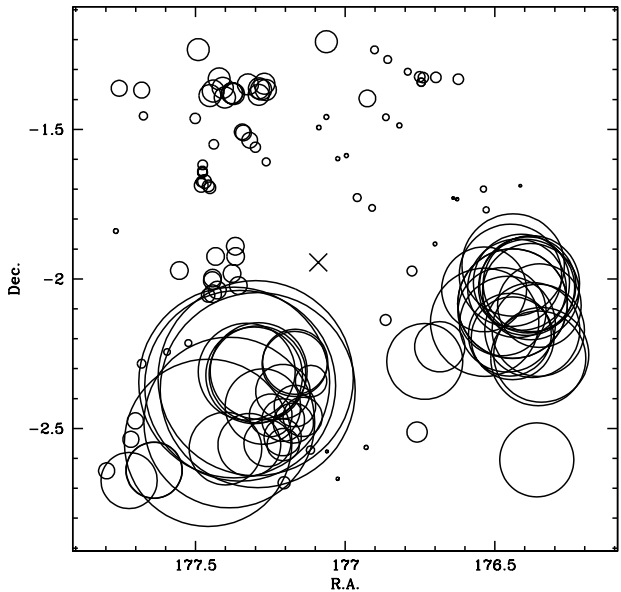,angle=0,width=3.4in}}
  \caption{DS test results of the first redshift peak 
identified in Table~\ref{tab:kmm1}.  
The cross denotes the position assigned to A1386 by Abell (1958).
Two major sub-clusters which contain BCMs ($\alpha$ and $\delta$) 
can be seen at (RA,Dec)=(177.3,$-$2.5) and (176.4,$-$2.1).
We suggest that these are two clusters
or groups in their own right.
}
\label{fig:dscone}
\end{figure}

In Fig.~\ref{fig:dscone}, we show the results of the DS test in
this redshift slice.  From this, it is quite clear that there
are two major self-contained sub-clusters (RA,Dec)=(177.3,$-$2.5) and
(176.4,$-$2.1) that dominate this slice.  
Moreover, these two sub-clusters, $\alpha$ and $\delta$, are among those
in which we identified typical BCMs.
It may also be the case
that the $\beta$ and $\gamma$ groups (Fig.~\ref{fig:groups})
are groups in their own right as well, but we have not been
able to identify dominant BCMs in these regions and they 
are probably false positives given the small number of
redshifts available (cf.\ Pinkney et al.\ 1996).  

The six-Gaussian solution to the KMM algorithm (lower half of
Table~\ref{tab:kmm1}) suggests that the three slices with the highest
redshift may also contain groups or clusters of galaxies in their own
right (indeed, we have already identified A1373 at 38063 kms$^{-1}$; 
see above).  
However, apart from A1373, the lack of BCMs coupled with no high galaxy
overdensity suggests otherwise.  The simplest interpretation
for these three redshift slices is that they are part of some
larger-scale structure.  Indeed, looking at the $cz>38000$ 
km s$^{-1}$ slice in Fig.~\ref{fig:groups} (which approximately 
corresponds with the highest redshift slice in Table~\ref{tab:kmm1})
the galaxies appear to be spread across the sky in the same
manner that walls and filaments of galaxies are 
(cf.\ Pimbblet et al.\ 2005).

To summarize, we present the global parameters for
the four groupings that we contend are bona-fide
clusters in their own right in Table~\ref{tab:new} from
all of the above analysis.  We estimate errors for 
$\sigma_v$ following Danese et al.\ (1980).
For the clusters that we name A1386-A and -B, we restrict our
computation of $\overline{cz}$ and $\sigma_v$ to the area of 
overlapping circles suggested by Fig.~\ref{fig:dscone} (i.e.\
coarsely splitting the $21000$--$27000$ kms$^{-1}$ redshift slice
at R.A.$=177$ and Dec.$<-1.9$)
and apply the clipping technique of Zabludoff et al.\ (1990).
For A1386-C and -D, we use all available redshifts in the
appropriate redshift slice (Figure~\ref{fig:groups})
over the full field of view (unlike for A1386-A and -B) and apply 
the same redshift clipping technique.
We also note that the clusters 
in Table~\ref{tab:new} extend beyond an Abell 
radius from the original Abell (1958) definition of A1386 -- 
this is apparent in the case of A1386-A where A1373 is the closest
companion in projection.  

%
% TABLE.  Bona Fide Clusters 
%
\begin{table*}
\begin{center}
\caption{Bona-fide clusters and groups with dominant BCMs in the field of A1386. 
N(gal) is the number
of galaxies used to derive $\overline{cz}$ and $\sigma_v$ and covers the
whole field of view for A1386-C and -D, but is spatially limited to the area of
the subclusters found in Fig.~7 for A1386-A and -B. \hfil}
\begin{tabular}{llllllrl}
\noalign{\medskip}
\hline
Name & N(gal) & $\overline{cz}$ &  $\sigma_v /$   & Other Cluster Names  & BCM & $v_{pec}$  & $v_{pec} / \sigma_v$ \\
     &        & (km\,s$^{-1}$)  &  (km\,s$^{-1}$) &              & (km\,s$^{-1}$)   & \\
\hline
A1386-A & 21 & $23593\pm68$ & $310^{+64}_{-39}$ & MZ 07066     & SDSS J114542.16$-$020151.9 ($\delta$) & $-$95 & $-$0.43 \\
A1386-B & 32 & $25952\pm83$ & $482^{+73}_{-50}$ & SDSS-C4 1121 & SDSS J114849.67$-$022817.6 ($\alpha$) & $-$297 & $-$0.88 \\ 
A1386-C & 56 & $28219\pm70$ & $451^{+60}_{-42}$ & MaxBCG J177.02469-01.68868 & SDSS J114805.92$-$014119.2 ($\epsilon$) & $-$61 & $-$0.15 \\
A1386-D & 66 & $30856\pm99$ & $818^{+81}_{-63}$ & Abell~1386 & SDSS J114814.35$-$015859.8 ($\zeta$; PRA037) & $+$108 & $+$0.15 \\
\hline
\noalign{\smallskip}
\end{tabular}
  \label{tab:new}
\end{center}
\end{table*}

Since our main, long-term aim in assembling these observations 
was to look more closely at peculiar velocities of BCMs
(cf. Coziol et al.\ 2009; Pimbblet 2008; Pimbblet et al.\ 2006),
we also compute the peculiar velocity, 
$v_{pec} = (v_{BCG} - \bar{v}_{cluster}) / (1+z)$, for
each group in Table~\ref{tab:new}.
Following Coziol et al.\ (2009), we express these
results as a fraction of the velocity dispersion, i.e.\ $v_{pec} / \sigma_v$,
to determine their level of significance (Table~\ref{tab:new}).

Two of the clusters in Table~\ref{tab:new} have BCM
peculiar velocities that are small fractions of the cluster velocity dispersions: 
A1386-C is $-0.15\sigma_v$ away from $\overline{cz}$; and A1386-D is $0.15\sigma_v$.  

Meanwhile, A1386-B is 
$-$0.88$\sigma_v$ away from $\overline{cz}$ and
A1386-A's BCM is $-$0.43$\sigma_v$ away from $\overline{cz}$.  
These values are large fractions of the velocity dispersion
(see Section 3.2 of Coziol et al. 2009 for a full discussion of
this parameter) and may be interesting groups 
for further observation.  
These peculiar velocity results are also in contrast to the values 
obtained by Coziol et al.\ (2009).  This is largely due to the
careful investigation of the A1386's substructure that we have
undertaken in this work and suggests that future investigation
of BCG peculiar velocities, especially in complex areas as this one,
must be undertaken with equal care in order to avoid erroneous outcomes.

As a final validation of these four groupings (Table~\ref{tab:new}), we construct
colour-magnitude diagrams for them and check whether we can see a defined 
colour-magnitude relation (e.g.\ Visvanathan \& Sandage 1977; Bower, Lucey \& Ellis 1992).
This is achieved by extracting SDSS ($g-r$) colours around the BCM of each group 
(Table~\ref{tab:new}) to a radius of 0.5$^{\circ}$ 
within $3 \sigma_v$ of the mean velocity given in Table~\ref{tab:new}
and comparing the location
of any obvious early-type ridge line with the 
empirical predictions of L{\'o}pez-Cruz, Barkhouse
\& Yee (2004; in particular see the equations contained in Figs.~3 and 4 of that work)
who surveyed clusters over a comparable redshift range to the present work.
One immediate issue in performing this analysis is that the equations 
presented by L{\'o}pez-Cruz et al.\ (2004) are in the Kron-Cousins
system (calibrated to Landolt 1992 standards), 
rather than SDSS $ugriz$ photometry.  Hence we transform the SDSS photometry to
Kron-Cousins using the transformations derived by Lupton\footnote{See 
http://www.sdss.org/dr4/algorithms/sdssUBVRITransform.html}.  Although Lupton 
derived these transformations for stellar objects, these transforms should hold for
galaxies that do not have significant emission lines -- just as one would  
expect for galaxies around the colour-magnitude relation.

\begin{figure*}
\vspace*{-3cm}
\centerline{\psfig{file=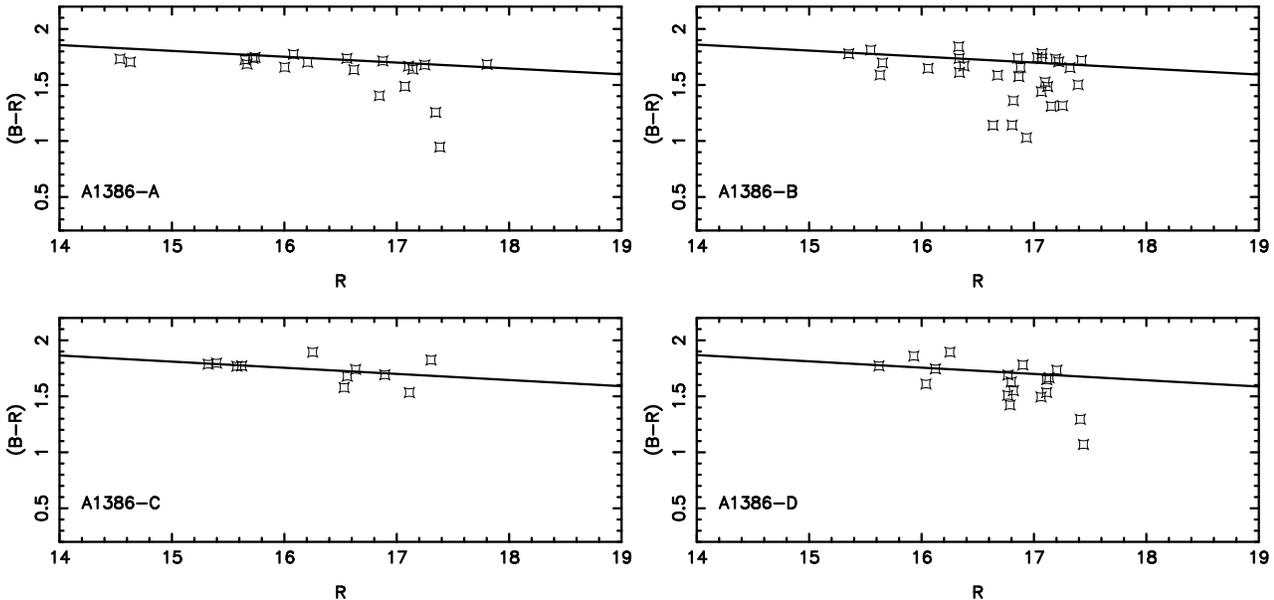,angle=0,width=7in}}
\vspace*{-12cm}
  \caption{Colour-magnitude diagrams for the groups identified in Table~\ref{tab:new}.  
The early-type ridge line is the empirical relation found 
by L{\'o}pez-Cruz et al.\ (2004). Each group has galaxies that are consistent 
with this relation, therefore adding weight to the finding that these groups
are real.
}
\label{fig:cmrs}
\end{figure*}

The resultant colour-magnitude diagrams are shown in Fig.~\ref{fig:cmrs} along
with the L{\'o}pez-Cruz et al.\ (2004) prediction for the early-type ridge line.
All of the four groupings show galaxies that are consistent with the 
predictions of where the ridge line should lie (especially given the errors
inherent in the photometric conversion to 
Kron-Cousins and the errors of the line given by L{\'o}pez-Cruz et al.\ 2004;
especially the scatter observed in Fig.~4 of L{\'o}pez-Cruz et al.\ 2004). 
Therefore we conclude that the field of A1386 is made up of 
at least four sub-units along the line of sight, and possibly a few filaments as
well, at the high redshift end of the distribution 
seen in Fig.~\ref{fig:zhist}.  We emphasize our observations and analysis of 
A1386 is wider than an Abell radius at these redshifts and therefore 
encompasses an area that Abell (1958) may not have examined in great detail.
In passing, we note that A1386-B seems to have a modest blue fraction 
of galaxies (e.g., Butcher \& Oemler 1984) indicating that it is more
highly star-forming than the other clusters we have detailed.

%
%  FIGURE.  Color-mag relations for the groups and clusters.
%

\section{Sloan Great Wall}

A1386 does not reside in just any part of the sky: it is 
a part of the Leo~A supercluster (Einasto et al.\ 1997)
and hence is a part of the Sloan Great Wall (SGW; Gott et al.\ 2005;
Nichol et al.\ 2006; Deng et al.\ 2007; Einasto et al.\ 2010).  
Although the SGW `may be the
largest coherent structure yet observed' (Tegmark et al.\ 2004) in the 
Universe, it could have still plausibly been formed from 
random phase Gaussian fluctuations (see Tegmark et al.\ 2004;
Gott et al.\ 2005) and even larger structures may yet be found 
(Shandarin 2009).  

\begin{figure*}
\centerline{\psfig{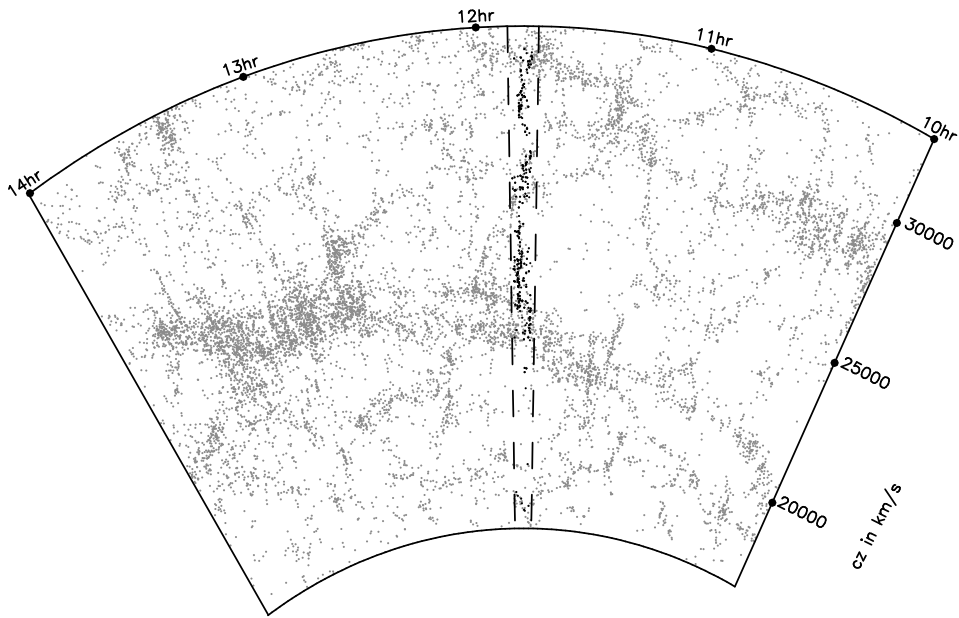}}
  \caption{Our new observations of A1386 combined with
all available literature redshifts
(black dots) plotted in the context
of the Sloan Great Wall (SDSS data shown as grey dots
in the declination range of
$1.5^{\circ}<\delta<-3.5^{\circ}$).  
The dashed lines denote the spatial extent of our AAOmega observations.
A1386 sits near the heart of the Great Wall in right ascension terms.
}
\label{fig:great}
\end{figure*}

In Fig.~\ref{fig:great} we demonstrate how our new radial velocities fit 
in with the SDSS measured velocities in the region of the SGW.
Our new measurements are probing the central regions of the 
SGW, at RA$\sim$11.8 hr.
This region of the SGW is where it has split into two defined filaments 
that stretch from $\sim11.3$ hr to $\sim12.7$ hr (cf.\ Fig.~9 of 
Gott et al.\ 2005).

%
%  FIGURE.  Great Wall Plot
%

The first point to make is that our new observations probe 
both the median $cz$ range of the SGW and beyond.  
At A1386's declination,
we qualitatively appear to be estabishing a 
previously hinted line-of-sight filament (Fig.~\ref{fig:great}).
This can readily be seen by contrasting the redshift histogram
(Fig.~\ref{fig:zhist}) with the points delineated by the dashed lines
in Fig.~\ref{fig:great} which denote the right ascension extent of our
observations.  

Our observations also better 
define the split in the SWG with higher fidelity than previous works.
Indeed, A1386-A resides at almost the Western end of this split, with A1386-B
residing beyond the two main filaments of the SGW.
Given the difference in $cz$ between A1386-B
(Table~\ref{tab:new}) and the bulk of the Great Wall, we suggest that the
interpretation is that this component 
A1386-B is likely to be infalling
along the line of sight to the `bulk' of the Great Wall at 
$cz\sim24000$~km s$^{-1}$.

This raises the question as to 
whether any or all of the four groups that we have 
identified (Table~\ref{tab:new}) are interacting with each other.  
The difference in recession velocity between the
groups is: 
$\Delta_{A to B}=2359$ km s$^{-1}$; 
$\Delta_{B to C}=2267$ km s$^{-1}$; 
$\Delta_{C to D}=2637$ km s$^{-1}$.
Given the small velocity dispersion of both A1386-A and A1386-B, it is
unlikely that these two groups are interacting.  
$\Delta_{B to C}$ is more than 5 times the velocity disperion of 
A1386-C ($\sigma_v=451$km s$^{-1}$), thus A1386-C is also unlikely to be 
interacting with A1386-B.
Even $\Delta_{C to D}$ is at least $>3.5\sigma_v$ of A1386-D.  
Hence we believe that
these groups and clusters are probably not interacting with each
other, although they may be moving coherently 
toward the bulk of the Great Wall.  
However, the analysis above does not preclude the possibility that
the groups found in the present work 
are connected by filaments extending in a radial direction
(cf.\ Lu et al.\ 2010) which could be verified with a deeper 
and more complete spectroscopic sample.  Indeed, Fig.~\ref{fig:great}
displays hints that all of our groups may be in proximity to
larger filaments (cf.\ Pimbblet et al.\ 2004).

\section{Summary}
This work has presented a catalogue of three hundred radial velocities 
(78 of them new) in the direction of A1386 as part of our 
on-going endeavours to
identify peculiar velocities of BCMs in Abell clusters. 
Here, we have taken a critical examination
of A1386 to unravel its complex architecture and its place
in the large-scale structure around it by looking at its relationship
with the Sloan Great Wall.  

We have demonstrated that 
A1386 is a not a simple, relaxed galaxy cluster.  It is composed of
at least four separate clusters, perhaps more, along the line of
sight that are spatially close to one another on the sky.  
We have delineated these structures through a combination of 
statistical substructure tests, presence of a dominant BCM,
and colour-magnitude relations.  The global parameters for these
new clusters are given in Table~\ref{tab:new}.  Of these, A1386-A and 
A1386-B both have 
a BCM with a peculiar velocity that is a large
fraction of their velocity dispersions.
Our results are in contrast to the 
results of Coziol et al.\ (2009), who report different subclusters,
and therefore different peculiar velocities.  The only common
subcluster between our two works is A1386A (as per Coziol et al.\ 2009),
which in our work has been divided in to A1386-C and A1386-D
(Table~\ref{tab:new}).  

There are several walls or filaments of galaxies that pass 
through the line of sight to A1386.  These are seen in Fig.~\ref{fig:zhist}
as peaks at $cz>33000$~km s$^{-1}$.  
A1386 is also located near the heart of the Sloan Great Wall.  
Of the sub-clusters identified, we suggest that only A1386-A is associated
with the SGW itself.
Although the other newly identified clusters may be moving toward other 
mass concentrations (both in and beyond the SGW), 
we suggest that at this time the other clusters are not
physically bound to the SGW and none of them are interacting with 
one another significantly.

\section*{Acknowledgements}
We would like to thank the dedicated staff at the 
Anglo-Australian Observatory for their support of our endeavours.  
We would like to explicitly thank Heath Jones 
for detailed discussion about 6dFGS data and Rob Sharp for advice about
AAOmega.
K.A.P.\ acknowledges partial support from the Australian Research Council.
H.A.\ has benefited from grants 50921-F, 81356, and 118295 of Mexican CONACyT.

This research has made use of the NASA/IPAC Extragalactic Database
(NED) which is operated by the Jet Propulsion Laboratory, California
Institute of Technology, under contract with the National Aeronautics
and Space Administration. 

We also acknowledge the usage of the HyperLeda database 
(http://leda.univ-lyon1.fr).

Funding for the SDSS and SDSS-II has been provided by the 
Alfred P. Sloan Foundation, the Participating Institutions, 
the National Science Foundation, the U.S. Department of Energy, 
the National Aeronautics and Space Administration, 
the Japanese Monbukagakusho, the Max Planck Society, and 
the Higher Education Funding Council for England.

The SDSS is managed by the Astrophysical Research Consortium for 
the Participating Institutions. The Participating Institutions are 
the American Museum of Natural History, Astrophysical Institute Potsdam, 
University of Basel, Cambridge University, Case Western Reserve University, 
University of Chicago, Drexel University, Fermilab, 
the Institute for Advanced Study, the Japan Participation Group, 
Johns Hopkins University, the Joint Institute for Nuclear Astrophysics, 
the Kavli Institute for Particle Astrophysics and Cosmology, 
the Korean Scientist Group, the Chinese Academy of Sciences (LAMOST), 
Los Alamos National Laboratory, the Max-Planck-Institute for Astronomy (MPIA), 
the Max-Planck-Institute for Astrophysics (MPA), 
New Mexico State University, Ohio State University, University of Pittsburgh, 
University of Portsmouth, Princeton University, 
the United States Naval Observatory, and the University of Washington.

\appendix
\section{Redshift catalogue}
In Table A1, we present the new radial velocities we measured from our AAOmega 
observations.  The designation `PRA' that appears in the table refers to 
three of the authors of this work (KAP, IGR \& HA) who created 
the original telescope time application, the target list and reduced the dataset.  

%\newpage
%
% TABLE.  Data Release.
%
\begin{table*}
\begin{center}
\caption{The redshift catalogue. The $R$ magnitude is sourced directly from the APM catalogue.
Where applicable, we have listed our targets that
also appear in other catalogues (right hand columns).
The sources for these are codified as follows:
(a) 2dFGRS (Colless et al. 2003);
(b) 6dFGS (Jones et al. 2009);
(c) HIPASS (Doyle et al.\ 2005);
(d) Quintana \& Ram\'\i rez (1995);
(e) LCRS (Shectman et al.\ 1996);
(f) UZC (Falco et al.\ 1999);
(g) SDSS-DR7 (Abazajian et al.\ 2009);
(h) Slinglend et al.\ (1998);
(i) Grogin et al.\ (1998);
(j) Da\,Costa L.N., et al.\ 1998.
%(k) 2QZ (Croom S.M. et al.\ 2004);
%(l) Theureau et al.\ (2005).
\hfil}
\begin{tabular}{lllrrrrrl}
\noalign{\medskip}
\hline
               &         &         &    &                        &            & \multicolumn{3}{c}{Other Measurements} \\
Identification & RA      & Dec     & $R$  & $cz_{helio}$           &  $\Delta cz$  & $cz_{helio}$   & $\Delta cz$  & \\
Tag            & (J2000) & (J2000) &    & (km\,s$^{-1}$) & (km\,s$^{-1}$)     & (km\,s$^{-1}$) & (km\,s$^{-1}$) & Source \\
\hline
PRA001 & 11 49 42.44 & $-$02 02 13.8 & 18.28 & 24720 & 101 & 24724 &  42 & g \\
PRA002 & 11 49 46.56 & $-$02 00 19.5 & 17.28 & 25527 & 128 & 25422 &  64 & a \\
       &             &               &       &       &     & 25347 &  49 & g \\
PRA003 & 11 49 48.83 & $-$01 47 27.2 & 15.64 &    26 &  29 &       &     & \\
PRA004 & 11 50 41.87 & $-$01 59 21.8 & 17.02 & 40714 & 110 & 40625 &  52 & g \\
PRA005 & 11 49 30.61 & $-$01 58 54.8 & 19.32 & 25545 &  20 &       &     & \\
PRA006 & 11 51 07.80 & $-$02 02 13.2 & 17.86 & 18161 &  74 & 17988 &  89 & a \\
       &             &               &       &       &     & 18072 &  48 & g \\
PRA007 & 11 50 53.60 & $-$02 03 07.4 & 19.22 & 80293 & 116 &       &     & \\
PRA008 & 11 49 25.68 & $-$02 01 19.7 & 16.75 & 25287 &  20 & 25237 &  43 & g \\
PRA009 & 11 49 56.40 & $-$02 05 23.8 & 16.63 & 34239 &  23 & 34188 &  42 & g \\
PRA010 & 11 50 46.70 & $-$02 01 47.8 & 19.18 & 48677 &  23 & 48536 & 123 & a \\
PRA011 & 11 50 12.00 & $-$02 04 19.2 & 19.46 & 37360 &  20 &       &     & \\
PRA012 & 11 51 11.97 & $-$02 05 31.7 & 17.25 & 48755 &  23 & 48776 &  64 & a \\
       &             &               &       &       &     & 48773 &  46 & g \\
PRA013 & 11 50 10.33 & $-$02 06 59.0 & 19.37 & 57800 &  23 &       &     & \\
PRA014 & 11 50 32.79 & $-$02 06 17.0 & 17.81 & 46911 &  23 & 46768 &  64 & a \\
       &             &               &       &       &     & 46777 &  53 & g \\
PRA015 & 11 50 04.83 & $-$02 07 56.2 & 19.13 & 34386 &  23 & 34326 &  89 & a \\
PRA016 & 11 49 46.28 & $-$02 02 58.6 & 18.62 & 25125 &  23 & 25123 &  89 & a \\
PRA017 & 11 51 10.20 & $-$02 07 00.5 & 19.14 & 67045 &  23 & 66914 &  89 & a \\
PRA018 & 11 50 05.65 & $-$02 12 51.8 & 18.15 & 26450 &  20 & 26382 &  89 & a \\
PRA019 & 11 50 33.54 & $-$02 13 35.3 & 17.93 & 40630 &  20 & 40592 &  64 & a \\
       &             &               &       &       &     & 40550 &  51 & g \\
PRA020 & 11 48 31.06 & $-$02 02 40.5 & 13.62 &  8493 &  17 &  8424 &  89 & a \\
       &             &               &       &       &     &  8502 &  28 & g \\
PRA021 & 11 48 59.41 & $-$02 00 31.0 & 15.92 &  8553 &  17 &  8574 &  64 & a \\
       &             &               &       &       &     &  8544 &  89 & b \\
       &             &               &       &       &     &  8574 &  27 & g \\
PRA022 & 11 49 33.56 & $-$02 04 36.8 & 18.92 & 43434 &  20 & 43500 &  89 & a \\
PRA023 & 11 50 26.64 & $-$02 10 18.4 & 16.21 & 18044 &  20 & 18027 &  23 & g \\
PRA024 & 11 48 08.66 & $-$02 07 46.9 & 18.96 & 44822 &  29 &       &     & \\
PRA025 & 11 48 32.97 & $-$02 00 18.5 & 17.75 & 30932 &  20 & 30930 &  25 & g \\
PRA026 & 11 48 50.40 & $-$02 01 56.0 & 10.93 &  1711 &  17 &  1704 &  45 & b \\
       &             &               &       &       &     &  1732 &   5 & c \\
       &             &               &       &       &     &  1736 & 100 & f \\
       &             &               &       &       &     &  1712 &   3 & g \\
PRA027 & 11 50 21.18 & $-$02 15 54.0 & 19.05 & 31972 &  20 & 31898 &  89 & a \\
PRA028 & 11 48 27.65 & $-$02 05 22.1 & 19.49 & 72507 &  23 &       &     & \\
PRA029 & 11 50 43.13 & $-$02 16 59.9 & 17.94 & 25533 & 131 & 25243 &  64 & a \\
       &             &               &       &       &     & 25288 &  45 & g \\
PRA030 & 11 50 14.01 & $-$02 28 07.3 & 16.80 & 27775 &  20 & 27761 &  64 & a \\
       &             &               &       &       &     & 27794 &  27 & g \\
PRA031 & 11 50 00.94 & $-$02 21 33.7 & 19.06 & 49753 &  29 & 49766 &  89 & a \\
PRA032 & 11 50 22.69 & $-$02 14 39.4 & 18.21 & 23320 &  23 & 23324 &  89 & a \\
PRA033 & 11 50 48.19 & $-$02 28 27.6 & 18.24 & 25239 &  20 & 25273 &  89 & a \\
PRA034 & 11 51 11.36 & $-$02 38 28.9 & 19.03 & 25266 &  23 &       &     & \\
PRA035 & 11 50 32.03 & $-$02 37 10.6 & 14.22 & 27760 &  20 & 27581 &  64 & a \\
       &             &               &       &       &     & 27629 &  64 & e \\
       &             &               &       &       &     & 27632 &  58 & g \\
PRA036 & 11 50 59.91 & $-$02 40 51.7 & 19.29 & 54607 &  26 &       &     & \\
PRA037 & 11 48 14.35 & $-$01 58 59.8 & 14.66 & 31022 &  23 & 30849 &  89 & a \\
       &             &               &       &       &     & 30930 &  45 & b \\
       &             &               &       &       &     & 30655 &  93 & d,1 \\
       &             &               &       &       &     & 30945 &  55 & g \\
\hline
\noalign{\smallskip}
\end{tabular}
\end{center}
\end{table*}

%\newpage
%
% TABLE.  Data Release.
%
\begin{table*}
\setcounter{table}{0}
\begin{center}
\caption{continued.
\hfil}
\begin{tabular}{lllrrrrrl}
\noalign{\medskip}
\hline
               &         &         &    &                        &            & \multicolumn{3}{c}{Other Measurements} \\
Identification & RA      & Dec     & $R$  & $cz_{helio}$           &  $\Delta cz$  & $cz_{helio}$   & $\Delta cz$  & \\
Tag            & (J2000) & (J2000) &    & (km\,s$^{-1}$) & (km\,s$^{-1}$)     & (km\,s$^{-1}$) & (km\,s$^{-1}$) & Source \\
\hline
PRA038 & 11 48 40.15 & $-$02 16 36.9 & 15.57 & 25482 &  20 & 25452 &  64 & a \\
       &             &               &       &       &     & 25440 &  39 & g \\
PRA039 & 11 48 35.99 & $-$02 19 23.0 & 19.05 & 39287 &  20 & 39273 & 123 & a \\
PRA040 & 11 49 29.08 & $-$02 20 46.2 & 18.77 & 26294 &  29 &       &     & \\
PRA041 & 11 49 24.16 & $-$02 21 24.8 & 18.29 & 25851 &  20 & 25572 &  89 & a \\
PRA042 & 11 49 32.87 & $-$02 28 48.5 & 15.99 & 26471 &  20 & 26262 &  89 & a \\
       &             &               &       &       &     & 26385 &  28 & g \\
PRA043 & 11 49 02.34 & $-$02 21 40.9 & 17.85 & 27404 &  20 & 27221 &  89 & a \\
       &             &               &       &       &     & 27398 &  28 & g \\
PRA044 & 11 50 23.07 & $-$02 41 19.6 & 19.21 & 13433 &  20 & 13461 &  64 & a \\
PRA045 & 11 48 49.84 & $-$02 22 34.8 & 17.11 & 26618 &  20 & 26592 &  89 & a \\
       &             &               &       &       &     & 26709 &  54 & g \\
PRA046 & 11 49 34.88 & $-$02 20 49.1 & 18.98 & 39326 &  20 & 39363 &  89 & a \\
PRA047 & 11 48 59.85 & $-$02 27 21.9 & 18.13 & 26114 & 119 & 26142 &  89 & a \\
       &             &               &       &       &     & 26028 &  48 & g \\
PRA048 & 11 48 57.70 & $-$02 31 00.4 & 17.25 & 30348 &  20 & 30099 &  89 & a \\
       &             &               &       &       &     & 30264 &  41 & g \\
PRA049 & 11 49 02.47 & $-$02 32 55.4 & 18.57 & 25980 & 110 & 25962 &  64 & a \\
       &             &               &       &       &     & 26109 &  44 & g \\
PRA050 & 11 49 12.96 & $-$02 19 20.1 & 17.31 & 26420 &  20 & 26622 &  64 & a \\
       &             &               &       &       &     & 26502 &  52 & g \\
PRA051 & 11 48 59.36 & $-$02 30 22.0 & 16.90 & 27865 &  20 & 27911 &  89 & a \\
       &             &               &       &       &     & 27827 &  48 & g \\
PRA052 & 11 48 56.75 & $-$02 31 53.9 & 15.07 & 17945 &  20 & 17928 &  89 & a \\
       &             &               &       &       &     & 17961 &  51 & g \\
PRA053 & 11 48 50.17 & $-$02 31 44.4 & 18.98 & 25446 &  65 & 25542 &  64 & a \\
PRA054 & 11 49 10.53 & $-$02 15 49.9 & 18.33 & 30437 &  20 & 30609 & 123 & a \\
PRA055 & 11 49 11.61 & $-$02 39 57.7 & 15.68 & 30551 &  68 & 30429 &  64 & a \\
       &             &               &       &       &     & 30532 &  45 & b \\
       &             &               &       &       &     & 30471 &  45 & g \\
       &             &               &       &       &     & 30523 &  43 & e \\
PRA056 & 11 49 05.91 & $-$02 25 35.1 & 17.04 & 25755 &  89 & 25782 &  64 & a \\
       &             &               &       &       &     & 25758 &  46 & g \\
PRA057 & 11 48 44.88 & $-$02 04 43.9 & 17.43 & 30878 &  20 & 30879 &  64 & a \\
       &             &               &       &       &     & 30903 &  38 & g \\
PRA058 & 11 49 18.38 & $-$02 34 20.5 & 17.58 & 27224 &  20 & 27311 &  89 & a \\
       &             &               &       &       &     & 27202 &  70 & e \\
       &             &               &       &       &     & 27212 &  30 & g \\
PRA059 & 11 49 18.93 & $-$02 18 35.4 & 16.98 & 26246 & 125 & 26262 &  64 & a \\
       &             &               &       &       &     & 26229 &  50 & g \\
PRA060 & 11 49 06.23 & $-$02 40 38.7 & 16.84 & 30168 &  20 & 30200 &  55 & e \\
       &             &               &       &       &     & 30207 &  25 & g \\
PRA061 & 11 47 28.42 & $-$02 04 09.0 & 18.36 & 34398 & 119 & 34206 &  64 & a \\
       &             &               &       &       &     & 34314 &  45 & g \\
PRA062 & 11 47 30.14 & $-$02 01 53.3 & 19.35 & 30740 &  23 &       &     & \\
PRA063 & 11 49 12.85 & $-$02 32 20.4 & 19.18 & 30812 &  20 & 30549 &  89 & a \\
PRA064 & 11 49 09.37 & $-$02 18 28.0 & 14.63 & 26618 &  20 & 26592 &  64 & a \\
       &             &               &       &       &     & 26619 &  51 & g \\
PRA065 & 11 47 26.57 & $-$02 05 36.4 & 19.05 & 47736 &  23 & 47877 &  89 & a \\
PRA066 & 11 48 41.50 & $-$02 28 54.5 & 18.37 & 25440 &  20 & 25392 &  89 & a \\
       &             &               &       &       &     & 25377 &  51 & g \\
PRA067 & 11 48 46.73 & $-$02 30 50.9 & 19.40 & 39314 &  20 &       &     & \\
PRA068 & 11 49 03.81 & $-$02 40 23.0 & 17.77 & 30381 &  20 & 30429 &  64 & a \\
       &             &               &       &       &     & 30432 &  30 & g \\
PRA069 & 11 47 34.66 & $-$02 08 20.9 & 17.92 & 17594 &  20 & 17598 &  89 & a \\
PRA070 & 11 48 30.63 & $-$02 18 20.2 & 17.72 & 30476 &  20 & 30399 &  64 & a \\
       &             &               &       &       &     & 30468 &  30 & g \\
PRA071 & 11 48 48.88 & $-$02 40 52.3 & 18.57 & 25308 &  65 & 25123 &  89 & a \\
       &             &               &       &       &     & 25285 &  49 & g \\
       &             &               &       &       &     & 25331 &  58 & e \\
PRA072 & 11 47 41.38 & $-$01 58 41.5 & 18.63 & 93100 &  26 &       &     & \\
\hline
\noalign{\smallskip}
\end{tabular}
\end{center}
\end{table*}

%\newpage
%
% TABLE.  Data Release.
%
\begin{table*}
\setcounter{table}{0}
\begin{center}
\caption{continued.
\hfil}
\begin{tabular}{lllrrrrrl}
\noalign{\medskip}
\hline
               &         &         &    &                        &            & \multicolumn{3}{c}{Other Measurements} \\
Identification & RA      & Dec     & $R$  & $cz_{helio}$           &  $\Delta cz$  & $cz_{helio}$   & $\Delta cz$  & \\
Tag            & (J2000) & (J2000) &    & (km\,s$^{-1}$) & (km\,s$^{-1}$)     & (km\,s$^{-1}$) & (km\,s$^{-1}$) & Source \\
\hline
PRA073 & 11 48 34.06 & $-$02 27 40.4 & 17.93 & 25653 &  20 & 25482 &  89 & a \\
PRA074 & 11 47 54.72 & $-$02 06 18.9 & 16.90 & 36790 &  20 & 36665 &  64 & a \\
       &             &               &       &       &     & 36758 &  16 & g \\
PRA075 & 11 48 40.42 & $-$02 25 29.4 & 17.35 & 25353 &  23 & 25333 &  89 & a \\
       &             &               &       &       &     & 25419 &  52 & g \\
PRA076 & 11 48 28.43 & $-$02 25 22.4 & 17.84 & 39434 &  62 & 39303 &  89 & a \\
       &             &               &       &       &     & 39381 &  42 & g \\
PRA077 & 11 47 39.86 & $-$02 23 42.8 & 17.26 & 38961 &  53 &       &     & \\
PRA078 & 11 47 44.69 & $-$02 24 19.1 & 19.42 & 50691 &  56 &       &     & \\
PRA079 & 11 47 43.02 & $-$02 15 23.8 & 18.16 & 39140 &  23 & 39243 &  89 & a \\
PRA080 & 11 47 45.13 & $-$01 57 06.5 & 15.82 & 34455 &  74 & 34386 & 123 & a \\
       &             &               &       &       &     & 34356 &  47 & g \\
PRA081 & 11 48 13.57 & $-$02 36 38.8 & 17.95 & 46045 &  80 & 45868 &  64 & a \\
       &             &               &       &       &     & 45961 &  49 & g \\
PRA082 & 11 47 43.39 & $-$02 30 48.9 & 19.19 & 31160 &  20 &       &     & \\
PRA083 & 11 48 32.03 & $-$02 26 36.2 & 18.47 & 30662 &  68 & 30339 &  89 & a \\
       &             &               &       &       &     & 30456 &  43 & g \\
PRA084 & 11 48 04.48 & $-$02 06 04.1 & 17.04 & 39827 &  23 & 39633 &  89 & a \\
       &             &               &       &       &     & 39947 &  48 & g \\
PRA085 & 11 47 43.13 & $-$02 33 47.7 & 17.30 & 23021 &  26 & 22994 &  64 & a \\
       &             &               &       &       &     & 23027 &  25 & g \\
PRA086 & 11 48 10.60 & $-$01 59 21.0 & 14.07 &  1555 &  17 &  1499 &  64 & a \\
       &             &               &       &       &     &  1529 &   5 & g \\
PRA087 & 11 48 27.35 & $-$02 20 29.3 & 18.30 & 25821 &  20 & 25692 &  89 & a \\
PRA088 & 11 48 04.11 & $-$01 57 59.0 & 18.13 & 31112 &  71 & 31089 &  64 & a \\
       &             &               &       &       &     & 31142 &  38 & g \\
PRA089 & 11 47 29.72 & $-$02 17 40.5 & 17.61 & 39116 &  65 & 38982 &  46 & g \\
PRA090 & 11 47 22.46 & $-$02 26 23.1 & 15.46 & 38445 &  20 & 38353 &  48 & g \\
       &             &               &       &       &     & 38463 &  64 & a \\
PRA091 & 11 47 17.24 & $-$02 23 35.5 & 19.35 & 39581 & 110 &       &     & \\
PRA092 & 11 48 29.70 & $-$02 21 48.8 & 17.82 & 30461 &  20 & 30549 &  89 & a \\
       &             &               &       &       &     & 30498 &  51 & g \\
PRA093 & 11 47 33.03 & $-$02 17 32.1 & 19.10 & 29835 & 122 & 29859 &  89 & a \\
PRA094 & 11 47 26.73 & $-$02 36 59.2 & 15.03 & 29628 &  95 & 29671 &  46 & g \\
PRA095 & 11 47 40.98 & $-$02 35 28.8 & 17.59 & 52562 &  89 & 52494 &  64 & a \\
       &             &               &       &       &     & 52383 &  58 & g \\
PRA096 & 11 47 23.25 & $-$02 22 31.9 & 18.11 & $-$44 &  83 &       &     & \\
PRA097 & 11 47 06.08 & $-$02 36 48.5 & 18.85 & 58513 &  23 &       &     & \\
PRA098 & 11 47 35.13 & $-$02 20 45.5 & 18.95 & 38535 &  20 & 38523 &  64 & a \\
PRA099 & 11 47 41.58 & $-$02 32 57.4 & 16.04 &  8556 &  32 &  8574 &  64 & a \\
       &             &               &       &       &     &  8511 &  54 & g \\
PRA100 & 11 46 52.86 & $-$02 34 41.5 & 17.82 & 28657 &  20 & 28678 &  33 & g \\
PRA101 & 11 47 10.86 & $-$02 10 26.1 & 13.94 & 17564 &  59 & 17652 &  23 & g \\
PRA102 & 11 47 39.82 & $-$02 31 31.8 & 16.28 &  8481 &  35 &  8424 &  64 & a \\
       &             &               &       &       &     &  8397 &  43 & g \\
PRA103 & 11 46 50.19 & $-$02 41 08.9 & 17.59 & 38592 &  26 & 39003 &  64 & a \\
       &             &               &       &       &     & 38967 &  41 & g \\
PRA104 & 11 46 47.17 & $-$02 22 57.6 & 19.27 & 39503 & 110 &       &     & \\
PRA105 & 11 47 15.50 & $-$02 01 47.4 & 19.28 & 34110 &  23 & 34116 & 123 & a \\
PRA106 & 11 47 09.47 & $-$02 31 58.3 & 16.27 & 13700 &  20 & 13692 &  25 & g \\
PRA107 & 11 47 03.57 & $-$02 22 06.5 & 17.57 & 39090 &  29 & 39093 &  64 & a \\
PRA108 & 11 46 54.91 & $-$02 32 16.8 & 15.53 &  8223 &  17 &  8244 &  64 & a \\
       &             &               &       &       &     &  8226 &  24 & g \\
       &             &               &       &       &     &  8196 &  86 & e \\
PRA109 & 11 47 21.45 & $-$02 00 07.7 & 15.92 & 34389 &  26 & 34266 &  64 & a \\
       &             &               &       &       &     & 34308 &  46 & g \\
PRA110 & 11 46 52.39 & $-$02 16 30.7 & 14.20 & 15547 &  20 & 15559 &  64 & a \\
       &             &               &       &       &     & 15594 &  45 & b \\
       &             &               &       &       &     & 15520 &  34 & g \\
\hline
\noalign{\smallskip}
\end{tabular}
\end{center}
\end{table*}

%\newpage
%
% TABLE.  Data Release.
%
\begin{table*}
\setcounter{table}{0}
\begin{center}
\caption{continued.
\hfil}
\begin{tabular}{lllrrrrrl}
\noalign{\medskip}
\hline
               &         &         &    &                        &            & \multicolumn{3}{c}{Other Measurements} \\
Identification & RA      & Dec     & $R$  & $cz_{helio}$           &  $\Delta cz$  & $cz_{helio}$   & $\Delta cz$  & \\
Tag            & (J2000) & (J2000) &    & (km\,s$^{-1}$) & (km\,s$^{-1}$)     & (km\,s$^{-1}$) & (km\,s$^{-1}$) & Source \\
\hline
PRA111 & 11 46 59.84 & $-$02 31 44.5 & 17.12 & 38700 &  23 & 38643 &  64 & a \\
       &             &               &       &       &     & 38694 &  42 & g \\
PRA112 & 11 46 30.81 & $-$02 33 13.9 & 15.44 & 13673 &  20 & 13641 &  64 & a \\
       &             &               &       &       &     & 13632 &  31 & g \\
PRA113 & 11 46 03.46 & $-$02 39 23.0 & 16.99 & 39344 &  20 & 39573 &  64 & a \\
       &             &               &       &       &     & 39525 &  44 & g \\
       &             &               &       &       &     & 39591 &  82 & e \\
PRA114 & 11 47 08.67 & $-$02 00 05.5 & 18.41 &  8475 &  17 &       &     & \\
PRA115 & 11 47 00.88 & $-$02 30 24.0 & 17.64 & 38583 &  20 & 38403 &  89 & a \\
       &             &               &       &       &     & 38496 &  23 & g \\
PRA116 & 11 46 06.49 & $-$02 41 36.0 & 17.99 & 33378 &  65 & 33316 &  40 & g \\
       &             &               &       &       &     & 33367 &  62 & e \\
PRA117 & 11 47 14.46 & $-$02 07 56.3 & 18.69 & 34287 &  20 &       &     & \\
PRA118 & 11 45 50.98 & $-$02 40 30.8 & 19.40 & 60081 &  26 &       &     & \\
PRA119 & 11 46 22.00 & $-$02 32 53.1 & 17.50 & 38442 &  74 & 38463 &  64 & a \\
       &             &               &       &       &     & 38394 &  40 & g \\
PRA120 & 11 46 10.02 & $-$02 26 19.1 & 17.84 & 36784 &  20 & 36785 &  64 & a \\
       &             &               &       &       &     & 36686 &  40 & g \\
PRA121 & 11 45 26.47 & $-$02 36 17.4 & 19.06 & 23245 &  23 & 23114 &  89 & a \\
PRA122 & 11 45 33.33 & $-$02 31 49.4 & 17.64 & 38469 &  23 & 38433 &  64 & a \\
       &             &               &       &       &     & 38388 & 270 & g \\
PRA123 & 11 45 30.37 & $-$02 37 21.7 & 15.62 & 15529 &  20 & 15469 &  64 & a \\
       &             &               &       &       &     & 15511 &  26 & g \\
       &             &               &       &       &     & 15511 &  40 & e \\
PRA124 & 11 45 37.05 & $-$02 25 35.8 & 16.69 & 34368 &  20 & 34356 &  64 & a \\
       &             &               &       &       &     & 34395 &  49 & g \\
PRA125 & 11 45 33.02 & $-$02 30 04.6 & 19.33 & 39569 &  20 &       &     & \\
PRA126 & 11 45 31.56 & $-$02 31 32.2 & 16.76 & 39374 &  98 & 39513 &  64 & a \\
       &             &               &       &       &     & 39450 &  51 & g \\
PRA127 & 11 46 08.80 & $-$02 17 50.2 & 17.13 & 36802 &  62 & 36605 & 123 & a \\
       &             &               &       &       &     & 36590 &  43 & g \\
PRA128 & 11 45 34.91 & $-$02 18 15.6 & 19.03 & 56735 &  23 &       &     & \\
PRA129 & 11 45 44.20 & $-$02 23 01.4 & 17.36 & 35153 &  56 & 35091 &  42 & g \\
PRA130 & 11 45 51.42 & $-$02 21 56.0 & 18.37 & 11625 &  77 & 36815 &  64 & a \\
       &             &               &       &       &     & 36851 &  42 & g \\
PRA131 & 11 45 43.34 & $-$02 19 47.9 & 17.26 & 35291 &  20 & 35436 &  64 & a \\
       &             &               &       &       &     & 35445 &  44 & g \\
PRA132 & 11 45 49.85 & $-$02 09 04.0 & 18.89 & 23473 &  20 &       &     & \\
PRA133 & 11 45 56.09 & $-$02 18 24.1 & 15.81 & 35900 & 101 & 35855 &  64 & a \\
       &             &               &       &       &     & 35867 &  45 & g \\
PRA134 & 11 45 45.88 & $-$02 11 54.3 & 17.56 & 23704 &  20 & 23594 & 123 & a \\
       &             &               &       &       &     & 23702 &  44 & g \\
PRA135 & 11 45 59.16 & $-$02 13 50.0 & 18.19 & 28357 & 188 & 28672 &  40 & g \\
PRA136 & 11 45 29.96 & $-$02 10 14.0 & 18.23 & 23905 &  20 & 23774 &  89 & a \\
PRA137 & 11 45 30.01 & $-$02 17 57.6 & 17.69 & 17720 &  20 & 17598 &  64 & a \\
       &             &               &       &       &     & 17721 &  24 & g \\
PRA138 & 11 46 03.16 & $-$02 14 45.7 & 18.94 & 36466 &  20 & 36395 &  64 & a \\
PRA139 & 11 45 25.30 & $-$02 15 55.9 & 16.76 & 24004 &  23 & 23968 &  44 & g \\
       &             &               &       &       &     & 23894 &  64 & a \\
PRA140 & 11 45 48.07 & $-$02 11 11.4 & 18.34 & 23923 &  20 & 23804 &  64 & a \\
PRA141 & 11 45 25.13 & $-$02 10 17.9 & 16.24 & 23944 & 188 & 23384 &  89 & a \\
       &             &               &       &       &     & 23375 &  52 & g \\
PRA142 & 11 46 18.57 & $-$02 12 54.3 & 18.07 & 34398 &  20 & 34506 &  64 & a \\
       &             &               &       &       &     & 34488 &  51 & g \\
PRA143 & 11 45 42.25 & $-$02 09 07.0 & 17.13 & 36448 &  23 & 36455 &  89 & a \\
       &             &               &       &       &     & 36460 &  50 & g \\
PRA144 & 11 45 24.28 & $-$02 05 21.8 & 17.06 & 23512 &  20 & 23468 &  26 & g \\
       &             &               &       &       &     & 23414 &  64 & a \\
PRA145 & 11 45 34.49 & $-$02 01 42.0 & 15.74 & 24430 & 203 & 23983 &  64 & a \\
       &             &               &       &       &     & 23968 &  52 & g \\
PRA146 & 11 45 28.96 & $-$02 03 29.6 & 18.90 & 61301 &  23 &       &     & \\
\hline
\noalign{\smallskip}
\end{tabular}
\end{center}
\end{table*}

%\newpage
%
% TABLE.  Data Release.
%
\begin{table*}
\setcounter{table}{0}
\begin{center}
\caption{continued.
\hfil}
\begin{tabular}{lllrrrrrl}
\noalign{\medskip}
\hline
               &         &         &    &                        &            & \multicolumn{3}{c}{Other Measurements} \\
Identification & RA      & Dec     & $R$  & $cz_{helio}$           &  $\Delta cz$  & $cz_{helio}$   & $\Delta cz$  & \\
Tag            & (J2000) & (J2000) &    & (km\,s$^{-1}$) & (km\,s$^{-1}$)     & (km\,s$^{-1}$) & (km\,s$^{-1}$) & Source \\
\hline
PRA147 & 11 45 56.52 & $-$02 08 39.0 & 19.29 & 60851 &  20 &       &     & \\
PRA148 & 11 47 12.24 & $-$01 53 51.5 & 18.92 & 28642 &  62 & 28630 &  64 & a \\
PRA149 & 11 46 31.34 & $-$01 55 28.5 & 17.92 & 36661 &  65 & 36755 &  64 & a \\
       &             &               &       &       &     & 36713 &  24 & g \\
PRA150 & 11 45 45.87 & $-$02 04 56.8 & 15.37 & 23683 & 329 & 23474 &  64 & a \\
       &             &               &       &       &     & 23303 &  55 & g \\
PRA151 & 11 47 38.44 & $-$01 55 22.1 & 18.24 &  8205 &  20 &       &     & \\
PRA152 & 11 46 27.86 & $-$01 53 56.1 & 19.09 & 36559 &  20 & 36275 & 123 & a \\
PRA153 & 11 46 33.66 & $-$01 58 22.4 & 18.92 & 31787 &  20 & 31748 &  64 & a \\
PRA154 & 11 46 21.51 & $-$01 55 11.8 & 16.09 & 36637 &  80 & 36485 &  64 & a \\
       &             &               &       &       &     & 36506 &  45 & g \\
PRA155 & 11 45 38.58 & $-$02 00 39.5 & 17.41 & 23731 &  26 & 23834 &  64 & a \\
       &             &               &       &       &     & 23923 &  43 & g \\
PRA156 & 11 46 24.28 & $-$01 53 24.5 & 18.08 & 14605 &  20 &       &     & \\
PRA157 & 11 45 45.47 & $-$01 57 12.8 & 15.29 & 23374 &  20 & 23234 &  64 & a \\
       &             &               &       &       &     & 23354 &  46 & g \\
PRA158 & 11 46 45.67 & $-$01 58 10.7 & 18.56 & 28171 & 131 & 28300 &  89 & a \\
PRA159 & 11 46 08.53 & $-$02 02 05.3 & 15.92 & 23081 & 314 & 23111 &  52 & g \\
PRA160 & 11 45 51.15 & $-$01 47 36.4 & 18.53 & 32044 &  20 &       &     & \\
PRA161 & 11 46 36.09 & $-$01 50 46.2 & 17.47 & 35000 &  23 &       &     & \\
PRA162 & 11 47 27.60 & $-$02 08 13.8 & 18.78 & 22112 &  20 & 22035 &  64 & a \\
PRA163 & 11 46 33.16 & $-$01 59 52.4 & 16.77 & 30428 &  20 & 30309 &  89 & a \\
       &             &               &       &       &     & 30447 &  54 & g \\
PRA164 & 11 46 47.91 & $-$01 52 58.5 & 14.79 & 22094 &  20 & 22095 &  89 & a \\
       &             &               &       &       &     & 22110 &  58 & g \\
PRA165 & 11 46 44.32 & $-$01 53 40.8 & 13.23 &  8385 &  23 &  8244 &  64 & a \\
       &             &               &       &       &     &  8238 &  52 & g \\
PRA166 & 11 45 22.41 & $-$01 52 36.2 & 11.68 &  8166 &  23 &  8124 &  64 & a \\
       &             &               &       &       &     &  8112 &  30 & f \\
       &             &               &       &       &     &  8091 &  46 & g \\
PRA167 & 11 45 40.00 & $-$01 48 58.4 & 18.32 & 32038 &  20 & 32078 &  64 & a \\
PRA168 & 11 46 40.77 & $-$01 49 39.0 & 17.30 & 36499 &  20 & 36485 &  64 & a \\
       &             &               &       &       &     & 36503 &  13 & g \\
PRA169 & 11 45 42.21 & $-$01 51 28.4 & 18.37 & 31948 &  20 &       &     & \\
PRA170 & 11 46 57.27 & $-$01 47 29.0 & 19.37 & 29742 &  20 &       &     & \\
PRA171 & 11 46 27.88 & $-$01 50 17.2 & 19.20 & 34587 &  20 &       &     & \\
PRA172 & 11 47 02.33 & $-$01 39 46.8 & 19.41 & 34131 &  20 &       &     & \\
PRA173 & 11 47 13.41 & $-$01 51 11.9 & 19.45 & 28849 &  20 &       &     & \\
PRA174 & 11 46 07.01 & $-$01 46 08.1 & 18.03 & 25185 &  20 & 25303 &  64 & a \\
       &             &               &       &       &     & 25201 &  53 & g \\
PRA175 & 11 48 03.32 & $-$01 40 02.1 & 19.26 & 28465 &  29 &       &     & \\
PRA176 & 11 47 06.50 & $-$01 58 24.8 & 18.59 & 24235 &  20 & 23594 &  64 & a \\
PRA177 & 11 46 34.49 & $-$01 40 51.5 & 18.24 &  8073 &  17 &       &     & \\
PRA178 & 11 46 59.08 & $-$01 58 18.0 & 16.48 & 30629 &  23 & 30579 &  64 & a \\
       &             &               &       &       &     & 30621 &  42 & g \\
PRA179 & 11 46 42.68 & $-$01 42 32.9 & 18.27 & 31882 &  68 & 31748 &  64 & a \\
       &             &               &       &       &     & 31835 &  38 & g \\
PRA180 & 11 46 03.96 & $-$01 47 02.4 & 16.42 & 19048 &  20 & 18977 &  64 & a \\
       &             &               &       &       &     & 19046 &  25 & g \\
PRA181 & 11 46 18.18 & $-$01 47 24.2 & 19.18 & 31897 &  20 &       &     & \\
PRA182 & 11 47 03.68 & $-$01 47 19.3 & 18.45 & 40765 &  20 & 40802 &  89 & a \\
PRA183 & 11 45 37.09 & $-$01 31 59.4 & 18.51 & 52970 & 134 & 52854 &  89 & a \\
PRA184 & 11 45 59.48 & $-$01 42 58.2 & 18.08 & 31778 & 110 & 31850 &  30 & g \\
PRA185 & 11 47 11.39 & $-$01 53 21.1 & 14.94 & 28186 & 146 & 28447 &  52 & g \\
PRA186 & 11 46 24.98 & $-$01 45 13.1 & 18.70 & 49432 &  23 & 49286 &  89 & a \\
PRA187 & 11 45 34.55 & $-$01 28 07.3 & 18.36 & 43209 & 143 &       &     & \\
PRA188 & 11 45 22.73 & $-$01 28 59.6 & 18.40 & 43077 &  23 &       &     & \\
PRA189 & 11 45 57.85 & $-$01 30 48.6 & 17.68 & 37150 &  71 & 36964 &  64 & a \\
       &             &               &       &       &     & 36955 &  39 & g \\
PRA190 & 11 46 54.07 & $-$01 52 33.8 & 17.13 & 29733 &  20 & 29829 &  89 & a \\
       &             &               &       &       &     & 29766 &  41 & g \\
\hline
\noalign{\smallskip}
\end{tabular}
\end{center}
\end{table*}

%\newpage
%
% TABLE.  Data Release.
%
\begin{table*}
\setcounter{table}{0}
\begin{center}
\caption{continued.
\hfil}
\begin{tabular}{lllrrrrrl}
\noalign{\medskip}
\hline
               &         &         &    &                        &            & \multicolumn{3}{c}{Other Measurements} \\
Identification & RA      & Dec     & $R$  & $cz_{helio}$           &  $\Delta cz$  & $cz_{helio}$   & $\Delta cz$  & \\
Tag            & (J2000) & (J2000) &    & (km\,s$^{-1}$) & (km\,s$^{-1}$)     & (km\,s$^{-1}$) & (km\,s$^{-1}$) & Source \\
\hline
PRA191 & 11 46 43.23 & $-$01 30 12.1 & 18.31 & 35120 &  80 & 34986 &  64 & a \\
       &             &               &       &       &     & 34980 &  42 & g \\
PRA192 & 11 45 33.33 & $-$01 24 34.5 & 17.19 & 35165 & 107 &       &     & \\
PRA193 & 11 45 49.51 & $-$01 28 50.8 & 16.30 & 28855 &  20 & 28960 &  89 & a \\
       &             &               &       &       &     & 28867 &  50 & g \\
PRA194 & 11 46 20.98 & $-$01 32 02.6 & 15.81 & 31685 &  95 & 31721 &  49 & g \\
PRA195 & 11 45 56.18 & $-$01 20 57.4 & 14.94 & 29007 &  23 & 28870 &  89 & a \\
PRA196 & 11 46 29.37 & $-$01 19 55.7 & 16.70 & 24184 &  20 & 24493 &  89 & a \\
PRA197 & 11 47 05.21 & $-$01 46 55.4 & 19.32 & 28423 &  20 &       &     & \\
PRA198 & 11 46 03.53 & $-$01 19 04.6 & 17.87 & 28810 &  20 & 28630 &  89 & a \\
PRA199 & 11 45 56.32 & $-$01 12 06.1 & 17.32 & 34689 &  20 & 34776 &  89 & a \\
       &             &               &       &       &     & 34831 &  20 & g \\
       &             &               &       &       &     & 35023 &  40 & h \\
PRA200 & 11 47 07.29 & $-$01 28 20.3 & 19.38 & 28600 &  20 &       &     & \\
PRA201 & 11 47 27.25 & $-$01 27 34.4 & 15.11 & 21977 &  20 &       &     & \\
PRA202 & 11 47 13.36 & $-$01 35 18.7 & 17.71 & 34182 &  20 & 34146 &  64 & a \\
       &             &               &       &       &     & 34170 &  44 & g \\
PRA203 & 11 46 36.74 & $-$01 21 17.2 & 17.44 & 27859 &  23 & 28061 &  89 & a \\
PRA204 & 11 46 36.21 & $-$01 16 23.3 & 18.96 & 27970 &  23 & 27971 &  89 & a \\
PRA205 & 11 46 58.96 & $-$01 20 31.9 & 17.53 & 24696 &  23 &       &     & \\
PRA206 & 11 46 47.30 & $-$01 19 34.1 & 17.96 & 25590 & 278 & 24583 &  64 & a \\
PRA207 & 11 47 03.41 & $-$01 19 11.2 & 16.24 & 31616 &  20 & 31598 &  64 & a \\
PRA208 & 11 46 51.64 & $-$01 29 06.1 & 16.33 & 27595 &  20 & 27371 &  89 & a \\
       &             &               &       &       &     & 27416 &  44 & g \\
PRA209 & 11 47 18.10 & $-$01 43 12.2 & 19.34 & 28039 &  23 &       &     & \\
PRA210 & 11 47 05.04 & $-$01 22 51.7 & 19.27 & 36041 &  26 &       &     & \\
PRA211 & 11 47 18.25 & $-$01 48 53.1 & 17.67 & 50545 &  23 & 50455 &  89 & a \\
       &             &               &       &       &     & 50566 &  50 & g \\
PRA212 & 11 47 38.61 & $-$01 21 21.3 & 17.73 & 32272 &  20 & 32018 & 123 & a \\
PRA213 & 11 47 13.41 & $-$01 27 13.7 & 19.18 & 28627 &  20 & 28450 &  64 & a \\
PRA214 & 11 47 41.20 & $-$01 18 01.9 & 19.46 & 79610 &  23 &       &     & \\
PRA215 & 11 47 16.58 & $-$01 29 12.2 & 16.35 & 23293 &  20 & 23354 &  89 & a \\
       &             &               &       &       &     & 23264 &  51 & g \\
PRA216 & 11 47 58.94 & $-$01 35 15.2 & 17.93 & 26699 &  26 & 26502 &  64 & a \\
       &             &               &       &       &     & 26628 &  42 & g \\
PRA217 & 11 47 23.03 & $-$01 45 23.7 & 14.11 & 34371 &  20 & 34356 &  89 & a \\
       &             &               &       &       &     & 34326 &  49 & g \\
PRA218 & 11 48 25.85 & $-$01 37 35.3 & 16.71 & 31433 &  23 & 31358 &  64 & a \\
PRA219 & 11 48 06.03 & $-$01 35 52.1 & 17.62 & 24897 &  20 & 24973 &  89 & a \\
       &             &               &       &       &     & 24892 &  21 & g \\
PRA220 & 11 48 06.02 & $-$01 33 35.9 & 16.48 & 28555 &  23 & 28540 &  64 & a \\
       &             &               &       &       &     & 28510 &  48 & g \\
PRA221 & 11 47 29.18 & $-$01 20 38.9 & 19.45 & 49570 &  23 &       &     & \\
PRA222 & 11 48 34.50 & $-$01 50 18.2 & 17.73 & 27853 & 164 & 28061 &  64 & a \\
       &             &               &       &       &     & 28217 &  46 & g \\
PRA223 & 11 48 02.19 & $-$01 37 04.3 & 19.28 & 43275 &  20 &       &     & \\
PRA224 & 11 47 50.50 & $-$01 43 40.7 & 17.69 & 24969 & 131 & 24793 &  64 & a \\
       &             &               &       &       &     & 24904 &  51 & g \\
PRA225 & 11 47 36.37 & $-$01 14 02.2 & 15.49 & 24034 &  20 & 24163 &  64 & a \\
PRA226 & 11 47 42.07 & $-$01 49 07.8 & 15.95 & 40846 &  23 & 40712 &  64 & a \\
       &             &               &       &       &     & 40847 &  45 & g \\
PRA227 & 11 48 20.70 & $-$01 56 39.5 & 18.95 & 31295 &  20 & 31508 &  89 & a \\
PRA228 & 11 47 51.20 & $-$01 56 06.2 & 15.30 & 34865 &  71 & 34776 &  64 & a \\
       &             &               &       &       &     & 34719 &  45 & g \\
PRA229 & 11 48 09.39 & $-$01 35 05.9 & 17.84 & 28576 &  23 & 28570 &  89 & a \\
       &             &               &       &       &     & 28543 &  43 & g \\
PRA230 & 11 47 38.39 & $-$01 45 44.9 & 18.52 & 25431 &  20 & 25392 &  64 & a \\
PRA231 & 11 48 39.92 & $-$01 12 07.7 & 18.30 & 31565 &  23 & 31478 &  64 & a \\
       &             &               &       &       &     & 31577 &  46 & g \\
PRA232 & 11 48 37.19 & $-$01 12 46.2 & 14.51 & 31999 &  59 & 31928 &  64 & a \\
       &             &               &       &       &     & 31991 &  48 & g \\
\hline
\noalign{\smallskip}
\end{tabular}
  \label{tab:cat}
\end{center}
\end{table*}

%\newpage
%
% TABLE.  Data Release.
%
\begin{table*}
\setcounter{table}{0}
\begin{center}
\caption{continued.
\hfil}
\begin{tabular}{lllrrrrrl}
\noalign{\medskip}
\hline
               &         &         &    &                        &            & \multicolumn{3}{c}{Other Measurements} \\
Identification & RA      & Dec     & $R$  & $cz_{helio}$           &  $\Delta cz$  & $cz_{helio}$   & $\Delta cz$  & \\
Tag            & (J2000) & (J2000) &    & (km\,s$^{-1}$) & (km\,s$^{-1}$)     & (km\,s$^{-1}$) & (km\,s$^{-1}$) & Source \\
\hline
PRA233 & 11 47 28.05 & $-$01 44 42.3 & 19.06 & 34101 &  20 & 34027 & 123 & a \\
PRA234 & 11 47 37.20 & $-$01 51 07.2 & 18.26 & 27565 & 146 & 27590 &  47 & g \\
PRA235 & 11 48 57.92 & $-$01 31 35.1 & 15.17 & 40903 &  26 & 40862 &  64 & a \\
       &             &               &       &       &     & 40832 &  49 & g \\
PRA236 & 11 48 15.08 & $-$01 27 30.5 & 19.02 & 24250 &  20 & 24193 &  64 & a \\
PRA237 & 11 48 11.54 & $-$01 40 54.7 & 19.21 & 27365 & 200 &       &     & \\
PRA238 & 11 48 07.04 & $-$01 42 58.4 & 19.01 & 27853 & 140 & 27821 & 123 & a \\
PRA239 & 11 48 40.45 & $-$01 42 54.8 & 17.84 & 36841 &  71 & 36665 &  64 & a \\
       &             &               &       &       &     & 36698 &  46 & g \\
PRA240 & 11 48 36.57 & $-$01 20 45.0 & 15.77 & 17795 &  20 &       &     & \\
PRA241 & 11 48 44.85 & $-$01 17 01.2 & 16.17 & 40660 &  20 &       &     & \\
PRA242 & 11 49 22.62 & $-$01 30 29.7 & 17.89 & 24819 & 122 & 24583 &  89 & a \\
       &             &               &       &       &     & 24667 &  51 & g \\
PRA243 & 11 48 21.08 & $-$01 50 48.4 & 18.96 & 30815 &  95 & 30639 & 123 & a \\
PRA244 & 11 48 35.46 & $-$01 31 49.0 & 15.54 &   155 &  62 &     0 &  64 & a \\
PRA245 & 11 47 44.10 & $-$01 51 15.5 & 16.28 & 30773 & 119 & 30759 &  64 & a \\
       &             &               &       &       &     & 30870 &  38 & g \\
PRA246 & 11 48 54.24 & $-$01 13 27.5 & 18.57 & 36769 &  71 & 36575 &  89 & a \\
       &             &               &       &       &     & 36563 &  43 & g \\
PRA247 & 11 49 03.29 & $-$01 36 31.3 & 17.03 & 24085 &  20 & 24082 &  26 & g \\
PRA248 & 11 48 21.21 & $-$01 29 37.4 & 15.66 & 24253 &  20 & 24073 &  64 & a \\
       &             &               &       &       &     & 24217 &  34 & g \\
PRA249 & 11 49 18.51 & $-$01 13 24.0 & 18.31 & 36026 &  20 & 36005 &  89 & a \\
PRA250 & 11 49 09.43 & $-$01 23 03.5 & 17.60 & 24523 & 116 & 24073 &  64 & a \\
PRA251 & 11 49 07.52 & $-$01 21 59.8 & 14.89 & 24262 &  20 & 24283 &  64 & a \\
PRA252 & 11 49 17.02 & $-$01 28 24.9 & 17.26 & 36961 & 149 & 36845 &  64 & a \\
       &             &               &       &       &     & 36934 &  40 & g \\
PRA253 & 11 49 41.01 & $-$01 13 49.6 & 17.87 & 37941 &  20 & 37564 &  89 & a \\
       &             &               &       &       &     & 37882 &  17 & g \\
PRA254 & 11 49 40.99 & $-$01 19 51.1 & 18.25 & 24412 &  20 & 24223 &  64 & a \\
PRA255 & 11 49 57.93 & $-$01 13 59.8 & 18.04 & 23998 &  20 & 24031 &  16 & g \\
PRA256 & 11 49 33.40 & $-$01 26 34.9 & 18.20 & 40708 & 110 & 40712 &  64 & a \\
PRA257 & 11 48 00.66 & $-$01 45 51.4 & 18.06 & 28540 &  20 &       &     & \\
PRA258 & 11 49 33.57 & $-$01 16 12.4 & 18.75 & 58972 &  23 & 58909 &  64 & a \\
PRA259 & 11 49 45.84 & $-$01 22 18.3 & 19.31 & 24322 &  20 &       &     & \\
PRA260 & 11 49 36.38 & $-$01 23 36.2 & 19.32 & 24693 &  20 &       &     & \\
PRA261 & 11 49 51.66 & $-$01 30 40.5 & 19.29 & 52682 &  29 &       &     & \\
PRA262 & 11 49 48.57 & $-$01 23 13.1 & 18.67 & 24058 &  32 &       &     & \\
PRA263 & 11 50 23.72 & $-$01 24 25.0 & 18.61 & 28012 & 107 & 28211 &  64 & a \\
PRA264 & 11 49 30.74 & $-$01 22 43.8 & 18.01 & 24942 & 296 & 24223 &  64 & a \\
PRA265 & 11 50 00.30 & $-$01 27 47.9 & 14.87 & 24100 &  23 & 24073 &  64 & a \\
       &             &               &       &       &     & 24085 &  49 & g \\
PRA266 & 11 50 27.38 & $-$01 26 15.1 & 19.17 & 37510 & 167 &       &     & \\
PRA267 & 11 49 36.35 & $-$01 27 19.9 & 10.93 &  5609 &  23 &  5563 &  45 & b \\
       &             &               &       &       &     &  5621 &  53 & g \\
       &             &               &       &       &     &  5634 &  34 & f \\
       &             &               &       &       &     &  5629 &  31 & i \\
       &             &               &       &       &     &  5612 &  32 & j \\
PRA268 & 11 50 18.51 & $-$01 20 34.5 & 18.85 & 39797 &  20 &       &     & \\
PRA269 & 11 50 33.21 & $-$01 21 15.9 & 14.56 & 48122 &  20 &       &     & \\
PRA270 & 11 49 12.01 & $-$01 33 34.5 & 17.80 & 24403 &  20 & 24313 &  64 & a \\
       &             &               &       &       &     & 24400 &  27 & g \\
PRA271 & 11 49 16.53 & $-$01 32 11.0 & 17.32 & 24753 &  20 & 24763 &  64 & a \\
       &             &               &       &       &     & 24724 &  23 & g \\
PRA272 & 11 51 01.17 & $-$01 21 45.2 & 18.64 & 24879 &  74 & 24703 &  64 & a \\
PRA273 & 11 51 09.29 & $-$01 24 44.3 & 18.97 & 48509 &  29 & 48536 &  89 & a \\
       &             &               &       &       &     & 49076 &  64 & a \\
PRA274 & 11 50 43.16 & $-$01 22 07.2 & 19.08 & 24864 &  20 &       &     & \\
PRA275 & 11 50 41.69 & $-$01 27 18.1 & 13.07 & 24130 &  20 & 24136 &  52 & g \\
       &             &               &       &       &     & 24135 &  45 & b \\
PRA276 & 11 50 10.49 & $-$01 28 45.8 & 19.14 & 40810 &  23 &       &     & \\
\hline
\noalign{\smallskip}
\end{tabular}
  \label{tab:cat}
\end{center}
\end{table*}

%\newpage
%
% TABLE.  Data Release.
%
\begin{table*}
\setcounter{table}{0}
\begin{center}
\caption{continued.
\hfil}
\begin{tabular}{lllrrrrrl}
\noalign{\medskip}
\hline
               &         &         &    &                        &            & \multicolumn{3}{c}{Other Measurements} \\
Identification & RA      & Dec     & $R$  & $cz_{helio}$           &  $\Delta cz$  & $cz_{helio}$   & $\Delta cz$  & \\
Tag            & (J2000) & (J2000) &    & (km\,s$^{-1}$) & (km\,s$^{-1}$)     & (km\,s$^{-1}$) & (km\,s$^{-1}$) & Source \\
\hline
PRA277 & 11 50 34.90 & $-$01 27 12.1 & 18.35 &  6091 &  20 &  5996 &  64 & a \\
PRA278 & 11 50 45.59 & $-$01 28 33.9 & 18.52 & 39818 &  20 &       &     & \\
PRA279 & 11 49 35.09 & $-$01 41 30.8 & 18.97 & 39761 & 131 & 39753 &  64 & a \\
PRA280 & 11 49 11.30 & $-$01 37 55.5 & 19.10 & 40070 &  20 & 40052 &  89 & a \\
PRA281 & 11 49 45.32 & $-$01 32 59.7 & 14.48 & 24873 &  20 & 24895 &  55 & g \\
PRA282 & 11 50 51.64 & $-$01 32 51.3 & 19.34 & 43565 &  20 &       &     & \\
PRA283 & 11 49 06.86 & $-$01 41 32.7 & 18.02 & 30285 &  20 & 30189 &  64 & a \\
PRA284 & 11 48 48.30 & $-$01 53 03.6 & 14.87 & 28234 &  23 & 28121 &  64 & a \\
       &             &               &       &       &     &  3788 &  55 & b,2 \\
       &             &               &       &       &     & 28232 &  51 & g \\
PRA285 & 11 49 14.18 & $-$01 38 45.4 & 18.32 & 36847 &  20 & 36845 &  64 & a \\
PRA286 & 11 49 59.05 & $-$01 47 17.2 & 18.65 & 37734 &  95 & 37384 & 123 & a \\
       &             &               &       &       &     & 37570 &  52 & g \\
PRA287 & 11 49 55.65 & $-$01 40 28.3 & 17.47 & 25119 &  23 & 25123 &  89 & a \\
       &             &               &       &       &     & 25147 &  45 & g \\
PRA288 & 11 49 55.50 & $-$01 41 13.3 & 18.41 & 25404 &  20 & 25333 &  89 & a \\
PRA289 & 11 49 54.21 & $-$01 37 06.7 & 18.41 & 25572 & 116 & 25243 &  64 & a \\
       &             &               &       &       &     & 25258 &  43 & g \\
PRA290 & 11 49 27.95 & $-$01 55 29.0 & 17.24 & 25434 & 104 & 25003 &  64 & a \\
       &             &               &       &       &     & 25207 &  46 & g \\
PRA291 & 11 48 45.51 & $-$01 51 08.5 & 19.47 & 34856 &  77 &       &     & \\
PRA292 & 11 49 32.91 & $-$01 48 17.9 & 17.74 & 30791 &  20 & 30849 &  64 & a \\
       &             &               &       &       &     & 30780 &  48 & g \\
PRA293 & 11 49 43.80 & $-$01 58 06.9 & 18.23 & 50578 &  29 & 50335 &  64 & a \\
       &             &               &       &       &     & 50413 &  54 & g \\
PRA294 & 11 49 34.42 & $-$01 39 16.0 & 17.03 & 52718 &  89 & 52494 &  64 & a \\
       &             &               &       &       &     & 52560 &  44 & g \\
PRA295 & 11 49 28.12 & $-$01 53 26.1 & 19.37 & 24846 &  20 &       &     & \\
PRA296 & 11 49 07.20 & $-$01 47 42.9 & 18.99 & 37998 &  20 & 37924 &  89 & a \\
PRA297 & 11 49 05.07 & $-$01 46 33.7 & 19.19 & 36496 & 161 &       &     & \\
PRA298 & 11 48 34.82 & $-$01 58 39.4 & 17.74 & 30369 &  23 & 30369 &  64 & a \\
       &             &               &       &       &     & 30411 &  43 & g \\
PRA299 & 11 49 20.86 & $-$01 50 59.2 & 17.73 & 39911 &  20 & 39813 &  64 & a \\
PRA300 & 11 50 46.61 & $-$01 57 36.7 & 13.54 &  5794 &  17 &  5795 &  54 & g \\
PRA301 & 11 49 48.49 & $-$01 41 42.3 & 18.51 & 25113 &  20 & 25063 &  64 & a \\
PRA302 & 11 51 11.47 & $-$01 51 26.4 & 19.21 & 28156 &  20 &       &     & \\
PRA303 & 11 50 13.03 & $-$01 58 18.7 & 16.62 & 25059 &  20 & 25081 &  35 & g \\
PRA304 & 11 49 36.75 & $-$01 48 45.8 & 19.46 & 52583 &  32 &       &     & \\
PRA305 & 11 50 33.32 & $-$02 01 51.7 & 17.28 & 40981 &  29 & 40892 &  64 & a \\
       &             &               &       &       &     & 40889 &  40 & g \\
PRA306 & 11 49 18.89 & $-$01 51 40.8 & 19.39 & 37483 &  20 &       &     & \\
PRA307 & 11 49 43.83 & $-$01 55 29.0 & 18.30 & 25161 &  68 & 24970 &  40 & g \\
\hline
\multicolumn{9}{l}{$^{a}$ Possibly a nucleus $\sim6$ arcsec SW of main galaxy?}\\
\multicolumn{9}{l}{$^{b}$ A poor quality 6dFGS spectrum (D.H.~Jones, priv. comm.)}\\
\noalign{\smallskip}
\end{tabular}
  \label{tab:cat}
\end{center}
\end{table*}

%\newpage
%\ \\
%\newpage
%\ \\
\newpage
\ \\
\newpage
\ \\
\newpage
%\ \\

\end{document}